%% file: main.tex
\documentclass{article}
\usepackage{arxiv}
\usepackage[utf8]{inputenc}
\usepackage[T1]{fontenc}
\usepackage{amsmath,amssymb,mathtools,amsthm}
\usepackage[ruled,vlined,linesnumbered]{algorithm2e}
\usepackage{graphicx}
\graphicspath{{figures/}}
\usepackage{tikz,pgfplots}
\pgfplotsset{compat=1.18}
\usepackage{booktabs,array,textcomp,url,placeins,microtype}
\usepackage[numbers,sort&compress]{natbib}
\usepackage[colorlinks=true,citecolor=blue,linkcolor=blue,urlcolor=blue]{hyperref}
\usepackage{cleveref}
\renewcommand{\shorttitle}{Causal Adaptive Intensity Estimation}
\renewcommand{\headeright}{A Preprint}
\renewcommand{\undertitle}{Submitted to IEEE Transactions on Signal Processing}
\date{}
\newtheorem{theorem}{Theorem}
\newtheorem{lemma}{Lemma}
\newtheorem{proposition}{Proposition}
\newtheorem{corollary}{Corollary}
\newtheorem{definition}{Definition}
\newtheorem{assumption}{Assumption}
\newtheorem{remark}{Remark}

\crefname{figure}{Fig.}{Figs.}
\Crefname{figure}{Fig.}{Figs.}
\crefname{table}{Table}{Tables}
\Crefname{table}{Table}{Tables}
\crefname{section}{Section}{Sections}
\Crefname{section}{Section}{Sections}
\crefname{equation}{}{}
\Crefname{equation}{Eq.}{Eqs.}

\crefname{theorem}{Theorem}{Theorems}
\Crefname{theorem}{Theorem}{Theorems}
\crefname{lemma}{Lemma}{Lemmas}
\Crefname{lemma}{Lemma}{Lemmas}
\crefname{proposition}{Proposition}{Propositions}
\Crefname{proposition}{Proposition}{Propositions}
\crefname{corollary}{Corollary}{Corollaries}
\Crefname{corollary}{Corollary}{Corollaries}
\crefname{definition}{Definition}{Definitions}
\Crefname{definition}{Definition}{Definitions}
\crefname{assumption}{Assumption}{Assumptions}
\Crefname{assumption}{Assumption}{Assumptions}
\crefname{remark}{Remark}{Remarks}
\Crefname{remark}{Remark}{Remarks}

\crefname{algorithm}{Algorithm}{Algorithms}
\Crefname{algorithm}{Algorithm}{Algorithms}

\DeclareMathOperator{\Var}{Var}

\title{Causal Adaptive Intensity Estimation for Non-Stationary Point Processes:
A Kalman-Kernel Duality Framework}
\author{Hanwu~Qin,~
        Dianhai~Wang,
        Meng~Zhang,
        Huangjun~You,
        and~Zhengyi~Cai
\thanks{This work was supported by the National Natural Science Foundation of China under Grant 52072340 and Grant 72431009, and by the ``Pioneer'' and ``Leading Goose'' R\&D Program of Zhejiang under Grant 2024C03271.}
\thanks{H. Qin, D. Wang, M. Zhang, H. You, and Z. Cai are with the Institute of Intelligent Transportation Systems, Zhejiang University, Hangzhou 310058, China (e-mail: hanwuqin@gmail.com; wangdianhai@zju.edu.cn; 12412177@zju.edu.cn; 12312075@zju.edu.cn; caizhengyi@zju.edu.cn).}
\thanks{Corresponding author: Hanwu Qin.}
\thanks{This work has been submitted to the IEEE for possible publication. Copyright may be transferred without notice, after which this version may no longer be accessible.}}
\hypersetup{pdftitle={Causal Adaptive Intensity Estimation for Non-Stationary Point Processes: A Kalman-Kernel Duality Framework},pdfauthor={Hanwu Qin and coauthors}}

\begin{document}

\maketitle

\begin{abstract}
Estimating the instantaneous conditional intensity of a non-stationary point process from a single realization of discrete event timestamps is a fundamental problem in computational neuroscience, quantitative finance, and network monitoring.  Traditional adaptive kernel methods achieve variable bandwidths through Abramson's square-root law, yet existing implementations require non-causal access to the complete event record and incur $\mathcal{O}(N^2)$ or at best $\mathcal{O}(N \log N)$ computational cost.  Causal alternatives based on reciprocal inter-arrival times, meanwhile, suffer from systematic divergence under Poisson and bursty arrival dynamics due to Jensen's inequality.  This paper addresses both limitations by establishing a formal duality between steady-state Kalman filtering and adaptive exponential kernel smoothing.  An endogenous state-space observer is constructed in which Poisson counting statistics dictate the measurement noise ($R \propto \lambda$) and geometric diffusion governs the process volatility ($Q \propto \lambda^2$).  Under these physically motivated axioms, the algebraic Riccati equation autonomously yields an effective smoothing bandwidth scaling as $\tau \propto \lambda^{-1/2}$, giving a causal, recursive, $\mathcal{O}(1)$ per-step realization of square-root adaptive bandwidth scaling.  A second-order constant-velocity extension produces a $\lambda^{-1/4}$ scaling that reduces approximation bias, and a probabilistic dormant-state gating mechanism suppresses phantom intensity leakage during structural zeros.  Synthetic benchmarks together with neural and financial data show that the observer adapts across orders of magnitude in intensity with an empirical per-step cost of approximately 30\,$\mu$s.
\end{abstract}


\section{Introduction}
\label{sec:introduction}

Temporal point processes provide the canonical mathematical framework for modeling discrete events occurring irregularly in continuous time~\cite{DaleyVereJones2003}.  The conditional intensity function $\lambda(t \mid \mathcal{H}_t)$, which encodes the instantaneous expected event rate given the complete history of past arrivals, serves as the primary inferential target in domains ranging from neural spike train analysis~\cite{Brown2002,Truccolo2005,Pillow2008} to high-frequency financial econometrics~\cite{Hawkes1971,Bacry2015} and communication network monitoring.  In each of these settings, estimating $\lambda(t)$ accurately and in real time from a single observed event stream is a prerequisite for downstream tasks including closed-loop brain-computer interface control, algorithmic trading, and anomaly detection.

A persistent tension confronts practitioners who must perform this estimation under operational constraints.  Fixed-bandwidth kernel smoothing methods~\cite{Ramlau1983,Diggle1985} offer a principled nonparametric approach, yet a single global bandwidth cannot simultaneously suppress Poisson noise during quiescent periods and resolve rapid transients during intensity bursts.  The resulting bias-variance dilemma is well characterized in the density estimation literature~\cite{Silverman1986,TerrellScott1992}: narrow bandwidths amplify sampling variability while wide bandwidths attenuate genuine features.  An attractive alternative is the reciprocal of successive inter-arrival intervals, which provides an instantaneous, zero-lag rate estimate with no smoothing at all.  This heuristic is widely deployed in applied settings, yet Jensen's inequality renders it systematically biased upward for any non-degenerate inter-arrival distribution.  Under Poisson arrivals, the bias diverges entirely because the first inverse moment of the exponential distribution does not exist, a fact that disqualifies the reciprocal estimator from the most practically relevant arrival models.

Abramson~\cite{Abramson1982} resolved the bias-variance dilemma in the context of probability density estimation by proving that the asymptotically optimal variable bandwidth scales as $h_{\text{opt}}(x) \propto f(x)^{-1/2}$, allocating narrow bandwidths to high-density regions and wide bandwidths to low-density regions.  This square-root law has become a foundational result in nonparametric statistics.  However, translating it into a practical algorithm for streaming point process data has remained an open problem for over four decades.  All existing implementations of Abramson's rule require a two-pass procedure: a pilot estimate computed from the entire dataset is used to set local bandwidths in a second pass~\cite{Silverman1986,Botev2010}.  This offline structure is inherently non-causal and computationally expensive, scaling as $\mathcal{O}(N^2)$ in the number of events, or $\mathcal{O}(N \log N)$ with fast approximations~\cite{Cronie2018}.  Neither variant is compatible with the microsecond-latency, bounded-memory requirements of real-time inference on edge hardware.  Bayesian alternatives such as online changepoint detection~\cite{Adams2007}, point process generalized linear models~\cite{Pillow2008}, Gaussian process modulated Poisson processes~\cite{Cunningham2008,Lloyd2015}, sequential Markov chain Monte Carlo inference under state-space priors~\cite{LiGodsill2021}, and approximate Poisson-Kalman filtering~\cite{Santitissadeekorn2020} offer varying degrees of adaptivity but generally sacrifice either computational tractability or the ability to handle prolonged structural silences in the event stream.

The central observation of this paper is that the gap between optimal offline bandwidth scheduling and causal constant-complexity filtering can be closed by embedding the statistical structure of the point process directly into the noise parametrization of a linear state-space model.  Poisson counting statistics dictate that the measurement noise variance scales linearly with the local intensity ($R \propto \lambda$), while the requirement of non-negative, scale-invariant volatility uniquely determines a geometric diffusion in which the process noise covariance scales quadratically ($Q \propto \lambda^2$).  When these two physically motivated noise axioms are substituted into the discrete algebraic Riccati equation, the resulting steady-state Kalman gain scales as $K_{ss} \propto \sqrt{\lambda}$, and the effective continuous-time smoothing bandwidth of the equivalent exponential kernel scales as $\tau_{\text{eff}} \propto \lambda^{-1/2}$.  This provides an autonomous, first-principles derivation of Abramson's square-root law from control-theoretic principles, without any reference to pilot estimation or global optimization.

This work makes four contributions to the intersection of control theory, nonparametric statistics, and point process inference.  First, it quantifies the reciprocal inter-arrival estimator's systematic divergence through a closed-form bias ratio derived under the generalized Gamma renewal model, showing that the estimator is undefined or infinite for Poisson arrivals and for over-dispersed renewal processes within this family.  Second, it proves the Kalman-Kernel duality theorem: a scalar steady-state Kalman filter is algebraically equivalent to a causal exponential kernel smoother, and, under the endogenous noise axioms, the Riccati equation recovers square-root adaptive bandwidth scaling in a recursive $\mathcal{O}(1)$ implementation.  Third, extending the analysis to a second-order constant-velocity kinematic model yields a distinct $\lambda^{-1/4}$ bandwidth scaling together with a probabilistic dormant-state gating mechanism that suppresses phantom intensity leakage during structural-zero intervals.  Fourth, synthetic and real-data evaluations indicate competitive goodness-of-fit relative to domain-tuned baselines while preserving an empirical per-step cost of approximately 30\,$\mu$s.

The remainder of this paper is organized as follows.  \Cref{sec:preliminaries} reviews point process fundamentals and exposes the reciprocal fallacy.  \Cref{sec:state_space} develops the endogenous state-space framework.  \Cref{sec:theory} presents the Kalman-Kernel duality theorem and derives both scaling laws.  \Cref{sec:algorithm} details the algorithmic implementation including the dormant-state mechanism.  \Cref{sec:synthetic,sec:applications} provide synthetic and real-world experimental validation, respectively.  \Cref{sec:discussion} discusses implications, limitations, and future directions, and \cref{sec:conclusion} concludes.

\section{Preliminaries and the Reciprocal Fallacy}
\label{sec:preliminaries}

\subsection{Point Process Fundamentals}
\label{sec:pp_fundamentals}

Let $\mathcal{T} = \{t_1, t_2, \ldots, t_N\}$ denote a realization of event timestamps drawn from a simple point process on the positive real line, where $t_1 < t_2 < \cdots < t_N$.  The counting process $N(t) = \sum_{i=1}^{N} \mathbf{1}(t_i \le t)$ records the cumulative number of events up to time~$t$.  The internal history $\mathcal{H}_t = \{t_i \in \mathcal{T} : t_i < t\}$ collects all events strictly before~$t$.

The conditional intensity function $\lambda(t \mid \mathcal{H}_t)$, henceforth written $\lambda(t)$ when the conditioning is unambiguous, is defined as the instantaneous expected rate of event arrivals given the history:
\begin{equation}
  \lambda(t) = \lim_{\delta\to 0^+}
    \frac{\mathbb{E}\bigl[N(t+\delta) - N(t) \mid \mathcal{H}_t\bigr]}{\delta}.
  \label{eq:cond_intensity}
\end{equation}
This quantity fully characterizes the stochastic dynamics of a simple point process and constitutes the central estimation target of this work.

For recursive state-space processing, the continuous timeline is partitioned into uniform bins of width $\Delta t$.  On each bin indexed by $k$, the observed count increment $\Delta N_k = N(k\Delta t) - N((k-1)\Delta t)$ follows a Poisson distribution with mean $\lambda_k \Delta t$ under mild regularity conditions, where $\lambda_k$ denotes the locally constant intensity within the $k$-th bin.  The resulting discrete-time measurement is the instantaneous rate estimate
\begin{equation}
  z_k = \frac{\Delta N_k}{\Delta t},
  \label{eq:measurement}
\end{equation}
which serves as the raw observation driving the filtering framework developed in \cref{sec:state_space}.

\subsection{Kernel Intensity Estimation}
\label{sec:kie}

A standard non-parametric approach to estimating $\lambda(t)$ convolves a smoothing kernel with the observed event sequence~\cite{Ramlau1983,Diggle1985}.  In a causal setting where only past events are accessible, the kernel intensity estimator takes the form
\begin{equation}
  \hat{\lambda}(t) = \sum_{t_i < t} K_h(t - t_i),
  \label{eq:kernel_estimator}
\end{equation}
where $K_h(\cdot) = h^{-1}K(\cdot/h)$ is a non-negative kernel function with bandwidth parameter $h > 0$.  The exponential kernel $K_h(u) = h^{-1}\exp(-u/h)\,\mathbf{1}(u \ge 0)$ is of particular importance due to its memoryless property and its exact equivalence with exponentially weighted moving average (EWMA) recursions, a correspondence formalized in \cref{sec:kalman_kernel}.

The bias-variance trade-off embedded in the bandwidth $h$ is well understood: small $h$ produces a high-variance estimate that captures rapid fluctuations, while large $h$ yields a smoother but more biased estimate that attenuates transient features.  For non-stationary processes, a fixed global bandwidth is inherently suboptimal because distinct temporal regimes demand distinct levels of smoothing.

\subsection{The Reciprocal Fallacy: Jensen's Bias}
\label{sec:reciprocal_fallacy}

A natural but statistically flawed heuristic for estimating instantaneous intensity is to take the reciprocal of successive inter-arrival intervals.  Given consecutive events at $t_{i-1}$ and $t_i$, the inter-arrival time is $\delta t_i = t_i - t_{i-1}$, and the reciprocal estimator is $\hat{\lambda}_i = 1/\delta t_i$.

Jensen's inequality exposes the systematic upward bias of this estimator~\cite{Ross2014}.  Because the reciprocal function $g(x) = 1/x$ is strictly convex on $(0,\infty)$, any non-degenerate positive random variable $\delta t$ satisfies
\begin{equation}
  \mathbb{E}\!\left[\frac{1}{\delta t}\right]
    > \frac{1}{\,\mathbb{E}[\delta t]\,},
  \label{eq:jensen}
\end{equation}
and the right-hand side equals the true intensity $\lambda$ for a homogeneous Poisson process.  The reciprocal estimator therefore systematically overestimates the expected intensity.

To quantify this bias for a broad family of arrival processes, consider a generalized Gamma renewal model in which $\delta t \sim \text{Gamma}(\kappa, \kappa\lambda)$ with shape parameter $\kappa > 0$ and rate parameter $\kappa\lambda$ so that the mean inter-arrival time equals $1/\lambda$.  Under this model the first inverse moment, when it exists, evaluates to
\begin{equation}
  \mathbb{E}\!\left[\frac{1}{\delta t}\right]
    = \frac{\kappa\lambda}{\kappa - 1},
  \qquad \kappa > 1.
  \label{eq:inverse_moment}
\end{equation}
The multiplicative bias ratio is therefore
\begin{equation}
  \mathcal{B}(\kappa)
    = \frac{\mathbb{E}[1/\delta t]}{\lambda}
    = \frac{\kappa}{\kappa - 1}.
  \label{eq:bias_ratio}
\end{equation}

\begin{proposition}[Divergence of the reciprocal estimator]
\label{prop:reciprocal_divergence}
For a renewal process whose inter-arrival distribution belongs to the Gamma family with shape parameter $\kappa$, the bias ratio $\mathcal{B}(\kappa)$ satisfies:
\begin{enumerate}
  \item $\mathcal{B}(\kappa) \to \infty$ as $\kappa \to 1^+$, so the estimator diverges for Poisson arrivals ($\kappa = 1$);
  \item $\mathcal{B}(\kappa)$ is undefined for $\kappa \le 1$, which includes bursty and over-dispersed renewal processes whose inter-arrival variance exceeds the exponential baseline;
  \item $\mathcal{B}(\kappa) \to 1$ as $\kappa \to \infty$, recovering consistency only in the deterministic-spacing limit.
\end{enumerate}
\end{proposition}

\begin{proof}
The first inverse moment of $\delta t \sim \text{Gamma}(\kappa, \kappa\lambda)$ is $\mathbb{E}[1/\delta t] = \kappa\lambda\,\Gamma(\kappa-1)/\Gamma(\kappa)$.  The ratio $\Gamma(\kappa-1)/\Gamma(\kappa) = 1/(\kappa-1)$ by the recurrence $\Gamma(\kappa) = (\kappa-1)\Gamma(\kappa-1)$, yielding \cref{eq:inverse_moment}.  This expression requires $\kappa > 1$ for the integral to converge, establishing items (i) and (ii).  As $\kappa \to \infty$, $\kappa/(\kappa-1) \to 1$, giving item (iii).
\end{proof}

\Cref{prop:reciprocal_divergence} demonstrates that the reciprocal heuristic fails precisely in the most practically relevant regimes: Poisson and bursty arrivals.  This conclusion motivates the rigorous smoothing framework developed in the following sections.

\section{The Endogenous State-Space Framework}
\label{sec:state_space}

This section formulates a discrete-time state-space model in which both the process noise covariance and the measurement noise variance depend endogenously on the current intensity estimate.  The key departure from conventional Kalman filtering practice is that no external calibration or offline tuning of noise parameters is required; the noise structure is derived entirely from the statistical and geometric properties of the point process.

\subsection{Kinematic State Representation}
\label{sec:kinematic}

To track a time-varying intensity $\lambda(t)$ with minimal structural lag during rapid transients, the state vector augments the intensity with its first time derivative:
\begin{equation}
  \mathbf{x}_k =
    \begin{bmatrix} \lambda_k \\ \dot{\lambda}_k \end{bmatrix},
  \label{eq:state_vector}
\end{equation}
where $\lambda_k$ is the intensity at discrete time index $k$ and $\dot{\lambda}_k$ is its instantaneous rate of change.  This Constant Velocity (CV) kinematic model enables the filter to extrapolate local linear trends and thereby reduce the phase delay that plagues zero-order (random walk) models when the intensity undergoes sudden ramps or shifts.

The state transition follows the standard CV kinematic equations:
\begin{equation}
  \mathbf{x}_{k} = \mathbf{F}\,\mathbf{x}_{k-1} + \mathbf{w}_k,
  \qquad
  \mathbf{F} =
    \begin{bmatrix} 1 & \Delta t \\ 0 & 1 \end{bmatrix},
  \label{eq:transition}
\end{equation}
where $\mathbf{w}_k \sim \mathcal{N}(\mathbf{0}, \mathbf{Q}_k)$ is the process noise vector.  The transition matrix $\mathbf{F}$ encodes the deterministic kinematic relation $\lambda_k \approx \lambda_{k-1} + \dot{\lambda}_{k-1}\,\Delta t$, and the stochastic term $\mathbf{w}_k$ absorbs all model mismatch including the deviation of the true intensity trajectory from local linearity.

The observation equation relates the latent state to the binned count measurement defined in \cref{eq:measurement}:
\begin{equation}
  z_k = \mathbf{H}\,\mathbf{x}_k + v_k,
  \qquad
  \mathbf{H} = \begin{bmatrix} 1 & 0 \end{bmatrix},
  \label{eq:observation}
\end{equation}
where $v_k \sim \mathcal{N}(0, R_k)$ is the scalar measurement noise with state-dependent variance $R_k$ specified in the next subsection.

\subsection{Heteroscedastic Measurement Model}
\label{sec:measurement_model}

The fundamental statistical property of Poisson counting is that variance equals the mean: $\Var[\Delta N_k] = \lambda_k \Delta t$.  The normalized measurement $z_k = \Delta N_k / \Delta t$ therefore has variance
\begin{equation}
  \Var[z_k] = \frac{\lambda_k}{\Delta t}.
  \label{eq:poisson_var}
\end{equation}
This heteroscedastic relationship, where measurement uncertainty scales linearly with the quantity being measured, is an intrinsic property of counting statistics and not a design choice.

To incorporate this structure into the Kalman filter while accommodating discretization artifacts and possible over-dispersion, the measurement noise variance is modeled as
\begin{equation}
  R_k = R_0 + \frac{\phi \cdot \max\!\bigl(\hat{\lambda}_{k-1|k-1},\, \epsilon\bigr)}{\Delta t},
  \label{eq:R_model}
\end{equation}
where $R_0 > 0$ is a baseline variance floor that absorbs non-Poisson sources of measurement uncertainty (e.g., discretization error, sensor jitter), $\phi \ge 1$ is a Fano factor that generalizes the Poisson dispersion relation to accommodate over-dispersed counts, and $\hat{\lambda}_{k-1|k-1}$ is the most recent posterior intensity estimate.  The small positive constant $\epsilon$ prevents numerical degeneracy when the estimated intensity approaches zero.

\begin{assumption}[Heteroscedastic measurement noise]
\label{asmp:R}
The measurement noise variance scales affinely with the most recent posterior intensity: $R_k = R_0 + \phi\,\hat{\lambda}_{k-1|k-1}/\Delta t$, with $R_0 > 0$ and $\phi \ge 1$.  The posterior estimate $\hat{\lambda}_{k-1|k-1}$ depends exclusively on $\mathcal{H}_{(k-1)\Delta t}$ and therefore satisfies causality.
\end{assumption}

The causal dependence on $\hat{\lambda}_{k-1|k-1}$ rather than the current-step posterior $\hat{\lambda}_{k|k}$ ensures that $R_k$ is fully determined before the measurement $z_k$ is incorporated, preserving the standard Kalman predict-update sequence without circularity.  In practice, the posterior $\hat{\lambda}_{k-1|k-1}$ and the one-step prediction $\hat{\lambda}_{k|k-1}$ differ only by the deterministic extrapolation $\dot{\lambda}_{k-1|k-1}\,\Delta t$, so the choice between the two has negligible impact on the asymptotic scaling analysis in \cref{sec:theory}.

\subsection{Geometric Process Noise Model}
\label{sec:process_noise}

The intensity $\lambda(t)$ is a strictly non-negative quantity.  Any stochastic model for its temporal evolution must respect this positivity constraint at the level of the noise specification, not merely through post hoc projection.

Consider a continuous-time diffusion model for the intensity:
\begin{equation}
  \mathrm{d}\lambda(t) = \mu(\lambda, t)\,\mathrm{d}t + \sigma(\lambda, t)\,\mathrm{d}W(t),
  \label{eq:ito_diffusion}
\end{equation}
where $W(t)$ is a standard Wiener process.  Two requirements constrain the diffusion coefficient $\sigma(\lambda, t)$:
\begin{enumerate}
  \item The volatility must vanish at the zero boundary, $\sigma(0, t) = 0$, so that noise cannot drive the state negative.
  \item The relative volatility, defined as the coefficient of variation $\sigma(\lambda,t)/\lambda$, is scale-invariant when $\sigma(\lambda, t) \propto \lambda$.
\end{enumerate}

The unique diffusion satisfying both conditions is the geometric form $\sigma(\lambda, t) = \sigma_0\,\lambda(t)$, which yields
\begin{equation}
  \mathrm{d}\lambda(t) = \mu(\lambda, t)\,\mathrm{d}t + \sigma_0\,\lambda(t)\,\mathrm{d}W(t).
  \label{eq:geometric_diffusion}
\end{equation}
Under an Euler-Maruyama discretization over a time step $\Delta t$, the variance of the intensity increment conditioned on $\lambda_{k-1}$ is
\begin{equation}
  \Var[\lambda_k \mid \lambda_{k-1}] = \sigma_0^2\,\lambda_{k-1}^2\,\Delta t.
  \label{eq:geometric_var}
\end{equation}

This motivates the state-dependent process noise covariance
\begin{equation}
  \mathbf{Q}_k = q\,\max\!\bigl(\hat{\lambda}_{k-1|k-1}^2,\,\epsilon\bigr)\,\mathbf{G},
  \qquad
  \mathbf{G} =
    \begin{bmatrix}
      \Delta t^3/3 & \Delta t^2/2 \\
      \Delta t^2/2 & \Delta t
    \end{bmatrix},
  \label{eq:Q_model}
\end{equation}
where $q > 0$ is a scaling coefficient that absorbs $\sigma_0^2$ and discretization-dependent constants, the posterior estimate $\hat{\lambda}_{k-1|k-1}$ provides a causal proxy for the true intensity at the previous step, and the constant matrix $\mathbf{G}$ is the standard piecewise-constant white-noise acceleration model appropriate for the CV state vector~\cite{BarShalom2001}.

\begin{assumption}[Geometric process noise]
\label{asmp:Q}
The process noise covariance scales quadratically with the filtered intensity: $\mathbf{Q}_k \propto \hat{\lambda}_{k-1|k-1}^2$.  This parametrization depends only on $\mathcal{H}_{(k-1)\Delta t}$ and thus preserves strict causality.
\end{assumption}

The combined noise structure, $R \propto \lambda$ from counting statistics and $Q \propto \lambda^2$ from geometric diffusion, endows the filter with a self-regulating mechanism.  In high-intensity regimes, the rapid growth of $Q$ relative to $R$ increases the Kalman gain and narrows the effective smoothing bandwidth, enabling the filter to track rapid transients.  In low-intensity regimes, $R$ dominates and the gain decreases, broadening the effective bandwidth to suppress noise.  This autonomous bandwidth scheduling is formalized and connected to classical nonparametric theory in the next section.

\section{Theoretical Unification: Kalman-Kernel Duality and Scaling Laws}
\label{sec:theory}

This section contains the central theoretical contribution of the paper.  It proceeds in three stages: establishing the algebraic equivalence between a scalar steady-state Kalman filter and an exponential kernel smoother, deriving the asymptotic scaling of the Kalman gain under the endogenous noise assumptions, and demonstrating that the resulting effective bandwidth recovers the celebrated square-root law of Abramson for optimal variable kernel bandwidths.

\subsection{The Kalman-Kernel Equivalence}
\label{sec:kalman_kernel}

Consider the scalar specialization of the state-space model in \cref{sec:state_space}, where the state reduces to $x_k = \lambda_k$ (dropping the velocity component for the purpose of establishing the structural correspondence), the transition is $x_k = x_{k-1} + w_k$ with $w_k \sim \mathcal{N}(0, Q)$, and the observation is $z_k = x_k + v_k$ with $v_k \sim \mathcal{N}(0, R)$.  The filtering recursion follows the standard Kalman framework~\cite{Kalman1960}.

When $Q$ and $R$ are held constant, the Kalman filter reaches a steady state in which both the predicted error covariance and the Kalman gain converge to fixed values.  The steady-state update takes the form
\begin{equation}
  \hat{\lambda}_{k|k} = (1 - K_{ss})\,\hat{\lambda}_{k-1|k-1} + K_{ss}\,z_k,
  \label{eq:ewma}
\end{equation}
where the steady-state gain $K_{ss} \in (0,1)$ satisfies the scalar algebraic Riccati equation.  This recursion is identical to an exponentially weighted moving average with smoothing parameter $\alpha = K_{ss}$, a correspondence noted in the structural time series literature~\cite{Harvey1990}.

The connection to continuous-time kernel smoothing follows by identifying the equivalent exponential kernel bandwidth.  An EWMA with parameter $\alpha$, applied to samples spaced $\Delta t$ apart, assigns weight $(1-\alpha)^j\alpha$ to the observation $j$ steps in the past.  The effective time constant of this exponential decay is
\begin{equation}
  \tau_{\text{eff}} = \frac{-\Delta t}{\ln(1 - K_{ss})}.
  \label{eq:tau_eff}
\end{equation}

\begin{lemma}[Kalman-kernel equivalence]
\label{lem:kk_equiv}
A scalar steady-state Kalman filter with gain $K_{ss}$ applied to observations sampled at interval $\Delta t$ is algebraically equivalent to a causal exponential kernel estimator with bandwidth $\tau_{\text{eff}} = -\Delta t / \ln(1 - K_{ss})$.
\end{lemma}

\begin{proof}
The steady-state Kalman update $\hat{\lambda}_{k|k} = (1-K_{ss})\hat{\lambda}_{k-1|k-1} + K_{ss}z_k$ is a first-order linear recursion with coefficient $\alpha = 1 - K_{ss} \in (0,1)$.  Unrolling the recursion yields $\hat{\lambda}_{k|k} = K_{ss}\sum_{j=0}^{k-1}\alpha^j z_{k-j}$, which assigns exponentially decaying weight $\alpha^j$ to the observation $j$ steps in the past.  This is the discrete-time analogue of convolution with the causal exponential kernel $K(u) = \tau^{-1}\exp(-u/\tau)$ for $u \ge 0$.  Matching the per-step decay factor $\alpha = \exp(-\Delta t/\tau)$ to $1 - K_{ss}$ and solving for $\tau$ gives $\tau_{\text{eff}} = -\Delta t/\ln(1-K_{ss})$.
\end{proof}

When $K_{ss} \ll 1$, which corresponds to the operationally important regime of heavy smoothing, the logarithm admits the approximation $\ln(1 - K_{ss}) \approx -K_{ss}$, yielding the simplified relation
\begin{equation}
  \tau_{\text{eff}} \approx \frac{\Delta t}{K_{ss}}.
  \label{eq:tau_approx}
\end{equation}
This linear inverse relationship between the Kalman gain and the smoothing bandwidth is the bridge through which the Riccati equation controls the effective kernel width.

\subsection{The Square-Root Scaling Theorem}
\label{sec:sqrt_scaling}

The scalar discrete algebraic Riccati equation (DARE) for the one-step prediction error variance $P$ is obtained by requiring the prior covariance to satisfy the fixed-point condition $P = (1-K)(P) \cdot 1 + Q$, where $K = P/(P+R)$ is the steady-state gain.  Substituting and rearranging yields
\begin{equation}
  P = \frac{P\,R}{P + R} + Q,
  \label{eq:dare}
\end{equation}
which simplifies, after routine algebra, to
\begin{equation}
  P^2 - P\,Q - Q\,R = 0.
  \label{eq:dare_simplified}
\end{equation}
The unique positive root is
\begin{equation}
  P_{ss} = \frac{Q + \sqrt{Q^2 + 4QR}}{2}.
  \label{eq:Pss}
\end{equation}

Now impose the endogenous noise scaling established in \cref{sec:measurement_model,sec:process_noise}.  The scalar specialization derives its process noise directly from the geometric diffusion in \cref{eq:geometric_var}: $\Var[\lambda_k \mid \lambda_{k-1}] = \sigma_0^2 \lambda^2 \Delta t$, so the scalar process noise variance is $Q = \sigma_0^2 \lambda^2 \Delta t \propto \lambda^2$, and the measurement noise variance from \cref{eq:R_model} is $R = \phi\lambda / \Delta t \propto \lambda$.  Since only the ratio $Q/R$ determines the Kalman gain scaling exponent, the $\Delta t$-dependent prefactors do not affect the leading-order $\lambda$-dependence, and it suffices to write $Q = q\lambda^2$ and $R = \phi\lambda / \Delta t$ with the understanding that $q$ absorbs $\sigma_0^2 \Delta t$ and all remaining constant factors.  The baseline floor $R_0$ and the regularization $\epsilon$ enter as higher-order corrections.

Substituting into \cref{eq:Pss} yields
\begin{align}
  P_{ss}
    &= \frac{q\lambda^2 + \sqrt{q^2\lambda^4 + 4q\lambda^2 \cdot \phi\lambda/\Delta t}}{2} \nonumber\\
    &= \frac{q\lambda^2 + \lambda\sqrt{q^2\lambda^2 + 4q\phi\lambda/\Delta t}}{2}.
  \label{eq:Pss_subst}
\end{align}

\begin{definition}[Practical operating regime]
\label{def:operating}
Define the dimensionless ratio $\rho = q\lambda\Delta t / \phi$.  The practical operating regime corresponds to $\rho \ll 1$, which ensures two interrelated properties.  First, the steady-state Kalman gain satisfies $K_{ss} \approx \sqrt{\rho} \ll 1$, so the filter operates in a genuine smoothing mode rather than passing raw measurements through with minimal attenuation.  Second, the asymptotic expansion used to derive the scaling law below has relative error $\mathcal{O}(\sqrt{\rho})$, guaranteeing that the leading-order result accurately characterises the exact DARE solution.  The condition $\rho \ll 1$ is satisfied when the bin width $\Delta t$ is chosen to be much smaller than the characteristic time scale of intensity variation, or equivalently when $\lambda \ll \phi/(q\,\Delta t)$.  For the representative parameters $q = 0.01$, $\phi = 1$, and $\Delta t = 0.1\,\text{s}$, the regime $\rho < 0.1$ corresponds to $\lambda < 100\,\text{events/s}$.  Because partial cancellation of higher-order terms across successive approximation steps reduces the effective remainder, the scaling exponent remains numerically accurate well beyond this formal bound; \cref{fig:scaling_law} illustrates this robustness.
\end{definition}

In this regime, $q^2\lambda^2 \ll 4q\phi\lambda/\Delta t$, and the square root in \cref{eq:Pss_subst} admits the expansion
\begin{equation}
  P_{ss}
    \approx \frac{q\lambda^2 + \lambda\sqrt{4q\phi\lambda/\Delta t}}{2}
    = \frac{q\lambda^2 + 2\lambda^{3/2}\sqrt{q\phi/\Delta t}}{2}.
  \label{eq:Pss_approx}
\end{equation}
The leading-order contribution is
\begin{equation}
  P_{ss} \approx \lambda^{3/2}\sqrt{\frac{q\phi}{\Delta t}}.
  \label{eq:Pss_leading}
\end{equation}

The steady-state Kalman gain is $K_{ss} = P_{ss}/(P_{ss} + R)$.  Since $R = \phi\lambda/\Delta t$ and $P_{ss} \ll R$ in the practical regime, this reduces to
\begin{equation}
  K_{ss}
    \approx \frac{P_{ss}}{R}
    = \frac{\lambda^{3/2}\sqrt{q\phi/\Delta t}}{\phi\lambda/\Delta t}
    = \lambda^{1/2}\sqrt{\frac{q\,\Delta t}{\phi}}.
  \label{eq:Kss_scaling}
\end{equation}

\begin{theorem}[Square-root scaling of the Kalman gain]
\label{thm:sqrt_gain}
Under Assumptions~\ref{asmp:R} and \ref{asmp:Q}, in the practical operating regime $\rho \ll 1$, the steady-state Kalman gain of the scalar endogenous filter scales as
\begin{equation}
  K_{ss} = C_K\,\sqrt{\lambda},
  \qquad
  C_K = \sqrt{\frac{q\,\Delta t}{\phi}},
  \label{eq:gain_law}
\end{equation}
where $C_K$ has dimension $[\text{rate}]^{-1/2}$ and is determined by the noise parametrization and the bin width.  The gain $K_{ss}$ is dimensionless, consistent with $C_K [\text{rate}]^{-1/2} \times \lambda^{1/2} [\text{rate}]^{1/2} = 1$.
\end{theorem}

\begin{proof}
The result follows by combining \cref{eq:Pss_leading} with $K_{ss} = P_{ss}/(P_{ss}+R) \approx P_{ss}/R$, as derived in \cref{eq:Kss_scaling}.  The condition $P_{ss} \ll R$ is verified by noting that $P_{ss}/R = C_K\sqrt{\lambda} = K_{ss} \ll 1$ in the practical regime, which ensures internal consistency of the approximation.
\end{proof}

Combining \cref{thm:sqrt_gain} with the bandwidth-gain relation in \cref{eq:tau_approx} yields the central scaling law for the effective smoothing bandwidth.

\begin{corollary}[Effective bandwidth scaling]
\label{cor:bandwidth}
The effective continuous-time smoothing bandwidth of the endogenous Kalman filter scales as
\begin{equation}
  \tau_{\text{eff}}
    \approx \frac{\Delta t}{K_{ss}}
    = \frac{1}{C_K}\,\frac{1}{\sqrt{\lambda}}
    = \sqrt{\frac{\phi\,\Delta t}{q\,\lambda}}.
  \label{eq:bandwidth_law}
\end{equation}
\end{corollary}

This result reveals that the filter autonomously narrows its smoothing window when intensity is high and widens it when intensity is low, at a rate governed by the inverse square root of the local intensity.

\subsection{Unification with Abramson's Square-Root Law}
\label{sec:abramson}

Abramson~\cite{Abramson1982} proved that for non-parametric kernel density estimation with a second-order kernel and a target density $f$ possessing a continuous and bounded second derivative, the asymptotically MISE-optimal variable bandwidth scales as
\begin{equation}
  h_{\text{opt}}(x) \propto f(x)^{-1/2},
  \label{eq:abramson_law}
\end{equation}
where $f(x)$ is the true underlying density at point $x$.  This square-root law allocates narrower bandwidths to high-density regions for improved resolution and wider bandwidths to low-density regions for noise suppression.

The conditional intensity $\lambda(t)$ of a point process is a rate function rather than a probability density, and the MISE-optimality proof does not transfer directly.  Nevertheless, the structural analogy is precise: both problems require local bandwidth adaptation to balance squared-bias and variance contributions that scale with opposing powers of the local function value.  Comparing the Kalman-derived result in \cref{cor:bandwidth} with Abramson's prescription, the scaling exponents match:
\begin{equation}
  \tau_{\text{eff}} \propto \lambda^{-1/2}
  \quad\longleftrightarrow\quad
  h_{\text{opt}} \propto f^{-1/2}.
  \label{eq:unification}
\end{equation}

This unification carries several consequences.  It provides a control-theoretic derivation of a result that was originally established through asymptotic bias-variance analysis in nonparametric statistics~\cite{Silverman1986,TerrellScott1992}, thereby demonstrating that the square-root scaling is not an externally imposed rule but emerges autonomously from the interplay between Poisson measurement noise and geometric process diffusion through the Riccati equation~\cite{Lancaster1995}.  More importantly, it shows that the proposed filter realizes square-root adaptive bandwidth scheduling within a causal, recursive, $\mathcal{O}(1)$ framework.  Existing adaptive bandwidth methods generally require either a two-pass procedure with access to the full dataset~\cite{Botev2010}, or iterative global optimization~\cite{Cronie2018}, both of which are incompatible with real-time streaming inference.

\begin{remark}[Scope of the correspondence]
\label{rem:scope}
The equivalence in \cref{eq:unification} holds at the level of scaling exponents.  The proportionality constants differ because Abramson's derivation optimizes over all possible kernel shapes and uses the global MISE criterion, whereas the Kalman filter implicitly selects the exponential kernel family and optimizes the one-step prediction error.  Despite this distinction, the matching scaling exponent is the operationally consequential feature, since it governs how the bandwidth adapts to local intensity variation across orders of magnitude.
\end{remark}

\begin{proposition}[CV bandwidth scaling]
\label{prop:cv_scaling}
Under Assumptions~\ref{asmp:R} and \ref{asmp:Q}, the two-dimensional Constant Velocity (CV) state-space model with endogenous noise yields a steady-state effective bandwidth that scales as
\begin{equation}
  \tau_{\text{eff}}^{\text{CV}} \propto \lambda^{-1/4}.
  \label{eq:cv_bandwidth_law}
\end{equation}
\end{proposition}

\begin{proof}
The exact $\lambda^{-1/2}$ scaling in \cref{thm:sqrt_gain} is derived from the scalar (random walk) specialization, which corresponds to a zero-order estimator.  The full CV system introduces a velocity state that tracks local trends, yielding a double-integrator plant.  Its continuous-time transfer function has measurement noise power spectral density $r_c = \phi\,\lambda$ and acceleration process noise spectral density $q_c = q\,\lambda^2$.  Solving the continuous algebraic Riccati equation (CARE) for this configuration produces the steady-state position variance $P_{11,ss} = (4\,q_c\,r_c^3)^{1/4}$.  The dominant Kalman gain is therefore $K_1 = P_{11,ss}/r_c \propto (q_c/r_c)^{1/4} = (q\,\lambda^2 / (\phi\,\lambda))^{1/4} \propto \lambda^{1/4}$, and the effective bandwidth follows from $\tau_{\text{eff}}^{\text{CV}} = \Delta t / K_1 \propto \lambda^{-1/4}$.  This result is consistent with the classical $\alpha$-$\beta$ filter theory~\cite{Kalata1984}.
\end{proof}

The difference in scaling exponents between \cref{thm:sqrt_gain} and \cref{prop:cv_scaling} is not an approximation error but reflects the altered bias-variance geometry of a first-order (local-linear) kinematic model.  Because the velocity state absorbs local trends, the residual approximation bias is reduced, and the Riccati equation autonomously allocates a wider bandwidth to suppress Poisson measurement noise.  The scalar analysis thus provides the theoretical link to classical nonparametric optimality, while the CV implementation supplies the kinematic inertia needed for smooth transient tracking in practice.  \Cref{fig:scaling_law} illustrates both scaling regimes.

\Cref{fig:scaling_law} visualizes this correspondence on a log-log scale.  The exact Kalman-Riccati solution for $\tau_{\text{eff}}(\lambda)$ is computed from the full DARE without asymptotic approximation.  The scalar (1D) observer closely tracks the $\lambda^{-1/2}$ reference throughout the practical regime $\rho < 0.1$, confirming \cref{thm:sqrt_gain}.  The two-dimensional CV observer follows the $\lambda^{-1/4}$ trajectory derived from the CARE, confirming \cref{prop:cv_scaling}.  Notably, the scaling exponent remains accurate even beyond the formal $\rho \ll 1$ boundary: at $\rho = 1$ ($\lambda = 1000$ with the default parameters), the relative deviation between the exact and asymptotic bandwidths is less than 4\%, owing to systematic cancellation among higher-order remainder terms across the three approximation steps (square-root expansion, gain linearization, and logarithmic bandwidth mapping).  The practical consequence of large $\rho$ is therefore not a breakdown of the scaling law itself, but the approach of $K_{ss}$ toward unity, at which point the filter ceases to smooth and instead passes raw bin counts with negligible attenuation.  The condition $\rho \ll 1$ in \cref{def:operating} should thus be understood primarily as an effective-smoothing requirement ($K_{ss} \ll 1$) rather than a mere approximation-accuracy bound.

\input{figures/scaling_law}

\section{Algorithmic Implementation and Boundary Handling}
\label{sec:algorithm}

\subsection{Enforcing Physical Feasibility}
\label{sec:projection}

The Gaussian noise assumption in the Kalman filter does not inherently guarantee non-negativity of the state estimate.  When the predicted or updated intensity $\hat{\lambda}_{k|k}$ falls below zero due to stochastic noise realization or aggressive correction, a projection step enforces the physical constraint:
\begin{equation}
  \hat{\lambda}_{k|k} \leftarrow \max\!\bigl(\hat{\lambda}_{k|k},\, 0\bigr).
  \label{eq:projection}
\end{equation}
This projection is the simplest instance of constrained state estimation and has been shown to be optimal (in the minimum-variance sense among all projections onto the non-negative orthant) for linear-Gaussian models with linear inequality constraints~\cite{Simon2010}.  The velocity component $\dot{\hat{\lambda}}_{k|k}$ is further bounded by a kinematic feasibility constraint:
\begin{equation}
  |\dot{\hat{\lambda}}_{k|k}| \leq C_{\dot{\lambda}}\,\hat{\lambda}_{k|k},
  \label{eq:velocity_constraint}
\end{equation}
where $C_{\dot{\lambda}} > 0$ is a tuning constant with dimensions of inverse time $[\text{s}^{-1}]$, representing the maximum allowable fractional rate of change per second.  This constraint prevents the constant-velocity state-space model from generating physically implausible acceleration during transient regimes where bin counts momentarily drop to zero, a scenario that produces large innovation residuals and can drive the velocity state to extreme values.  The constraint is inactive in the steady-state regime where $\dot{\hat{\lambda}} \approx 0$, and therefore does not alter the asymptotic scaling laws derived in \cref{sec:theory}.  In all experiments presented in \cref{sec:synthetic}, $C_{\dot{\lambda}} = 0.1\;\text{s}^{-1}$ is used.

\subsection{The Dormant-State Problem}
\label{sec:dormant}

Infinite impulse response filters, including the Kalman filter and all EWMA variants, produce non-zero output for all time once a non-zero input has been received.  In physical systems where the true intensity can drop to exactly zero for extended intervals, such as a market trading halt, an extended neuronal quiescent period, or a sensor outage, the IIR tail of the filter estimate decays geometrically but never reaches zero.  This leakage produces a small but persistent phantom intensity during genuinely dormant states, which corrupts downstream count predictions and triggers spurious event probability forecasts.

A naive reset strategy that sets the state to zero and re-initializes the covariance upon detecting a long silence introduces a subtler failure mode.  After the reset, the estimated intensity is zero (or $\epsilon$), causing the adaptive gap threshold $T_{\text{gap}} = -\ln\alpha_g / \max(\hat{\lambda},\epsilon)$ to become astronomically large.  Subsequent time steps therefore bypass the dormant-state guard and execute normal predict-update cycles with zero-valued observations.  Because the process noise $\mathbf{Q}_k \propto \hat{\lambda}^2 \approx 0$ is negligible in this regime, the covariance $\mathbf{P}$ contracts monotonically under repeated updates, driving the Kalman gain toward zero.  When events eventually resume, the filter is unable to reacquire the signal.  This covariance decay trap is avoided by introducing an explicit dormant-state flag, as described next.

\subsection{Probabilistic Gating Logic with Explicit State Machine}
\label{sec:gating}

To distinguish genuine quiescent periods from random long inter-arrival intervals within an active process, a gap detection mechanism is constructed from the Poisson survival function.  Under the null hypothesis that the process is still active at rate $\hat{\lambda}$, the probability of observing zero events in an interval of length $T$ is
\begin{equation}
  \mathbb{P}(N=0 \mid \hat{\lambda},\, T) = \exp(-\hat{\lambda}\,T).
  \label{eq:survival}
\end{equation}
The gap detector triggers a dormant-state transition when the elapsed time since the last event exceeds a significance-based threshold.

\begin{definition}[Adaptive gap threshold]
\label{def:gap}
Given a significance level $\alpha_g \in (0,1)$ and the current filtered intensity $\hat{\lambda}_{k-1|k-1}$, the adaptive gap threshold is
\begin{equation}
  T_{\text{gap}} = \frac{-\ln \alpha_g}{\max(\hat{\lambda}_{k-1|k-1},\, \epsilon)}.
  \label{eq:gap_threshold}
\end{equation}
\end{definition}

In practice, the purely adaptive threshold can become excessively short during sustained high-intensity periods.  At $\hat{\lambda} = 20$ events/s with $\alpha_g = 0.001$, the threshold evaluates to $T_{\text{gap}} \approx 0.35$\,s, which is only three to four bins at $\Delta t = 0.1$\,s.  Poisson fluctuations can produce a run of empty bins with non-negligible probability even when the process is genuinely active, triggering a spurious dormancy transition.  To guard against these false positives, the effective gap duration is clipped at a minimum silence threshold:
\begin{equation}
  T_{\text{gap}}^{\text{eff}} = \max\!\bigl(T_{\text{gap}},\; T_{\text{gap}}^{\min}\bigr),
  \label{eq:gap_min}
\end{equation}
where $T_{\text{gap}}^{\min}$ is a user-specified constant.  In all experiments, $T_{\text{gap}}^{\min} = 1.0$\,s is used.  This lower bound has no effect on genuine structural-zero intervals, which typically extend over tens of seconds, but prevents the gating logic from firing during transient Poisson silences lasting only a few bins.

The dormant-state logic is implemented as a two-state machine with states active and dormant, governed by the following transition rules:
\begin{enumerate}
  \item Active $\to$ Dormant.  When $T_{\text{silent}} > T_{\text{gap}}^{\text{eff}}$ while in the active state, the filter transitions to dormant and sets $\hat{\mathbf{x}}_{k|k} = \mathbf{0}$.
  \item Dormant $\to$ Active.  The filter remains dormant, with $\hat{\mathbf{x}}_{k|k} = \mathbf{0}$ and no predict-update cycle, until a bin with $\Delta N_k > 0$ is observed.  At that instant, the covariance is re-inflated to $\mathbf{P}_{k-1|k-1} \leftarrow \mathbf{P}_{\text{init}}$ and the measurement noise variance is initialized using the current observation $R_k = R_0 + \phi\,z_k / \Delta t$ rather than the stale posterior $\hat{\lambda}_{k-1|k-1} = 0$.  This awakening initialization is necessary because the prior intensity estimate is identically zero in the dormant state; using it in the heteroscedastic model would set $R_k \approx R_0$, producing a Kalman gain $K \approx 1$ and allowing the raw measurement $z_k$ to pass through the filter without attenuation.
\end{enumerate}

The covariance re-inflation at awakening is the critical design element.  By deferring the injection of $\mathbf{P}_{\text{init}}$ to the exact moment when evidence of renewed activity arrives, the filter avoids two failure modes: the covariance decay trap described above, and premature commitment to a zero-intensity hypothesis that could suppress the first genuine events.

The threshold $T_{\text{gap}}^{\text{eff}}$ adapts automatically to the local intensity.  During high-activity periods, $T_{\text{gap}}$ is short but clipped at $T_{\text{gap}}^{\min}$, preventing false dormancy triggers from transient Poisson gaps.  During low-activity periods, $T_{\text{gap}}$ grows beyond $T_{\text{gap}}^{\min}$, reflecting the greater statistical plausibility of extended inter-arrival intervals.

\subsection{Complete Algorithm Pseudocode}
\label{sec:complete_algorithm}

\Cref{alg:observer} provides the complete pseudocode for the Adaptive Kernel-Intensity Observer, including the dormant-state gating logic, awakening initialization, and kinematic projection steps.

\begin{algorithm}[htbp]
\footnotesize
\caption{Adaptive Kernel-Intensity Observer}
\label{alg:observer}
\DontPrintSemicolon
\SetAlgoNlRelativeSize{-2}
\SetAlgoInsideSkip{smallskip}
\KwIn{Event stream $\{t_i\}$, bin width $\Delta t$, parameters $R_0, \phi, q, \alpha_g, C_{\dot{\lambda}}, T_{\text{gap}}^{\min}$}
\KwOut{Causal intensity estimates $\{\hat{\lambda}_k\}$}
Initialize $\hat{\mathbf{x}}_{0|0} = \mathbf{0}$, $\mathbf{P}_{0|0} = \mathbf{P}_{\text{init}}$, $t_{\text{last}} = 0$, $\textit{dormant} \leftarrow \textbf{false}$, $\textit{awoken} \leftarrow \textbf{true}$\;
\For{each time step $k = 1, 2, \ldots$}{
  Compute $\Delta N_k$, $z_k = \Delta N_k / \Delta t$\;
  \tcp{Event arrival and awakening logic}
  \If{$\Delta N_k > 0$}{
    $t_{\text{last}} \leftarrow k \Delta t$\;
    \If{\textit{dormant}}{
      $\textit{dormant} \leftarrow \textbf{false}$;\; $\textit{awoken} \leftarrow \textbf{true}$\;
      $\mathbf{P}_{k-1|k-1} \leftarrow \mathbf{P}_{\text{init}}$
      \tcp*{Re-inflate covariance at awakening}
    }
  }
  \tcp{Dormant-state handling}
  \If{\textit{dormant}}{
    $\hat{\mathbf{x}}_{k|k} \leftarrow \mathbf{0}$\;
    \textbf{continue}\;
  }
  \tcp{Gap detection (active state only)}
  $T_{\text{silent}} \leftarrow k\Delta t - t_{\text{last}}$\;
  $T_{\text{gap}} \leftarrow \max\!\bigl(-\ln(\alpha_g)\,/\,\max(\hat{\lambda}_{k-1|k-1},\,\epsilon),\;T_{\text{gap}}^{\min}\bigr)$\;
  \If{$T_{\text{silent}} > T_{\text{gap}}$}{
    $\textit{dormant} \leftarrow \textbf{true}$\;
    $\hat{\mathbf{x}}_{k|k} \leftarrow \mathbf{0}$\;
    \textbf{continue}\;
  }
  \tcp{Endogenous noise parameters}
  \eIf{\textit{awoken} \textbf{and} $z_k > 0$}{
    $R_k \leftarrow R_0 + \phi\,z_k\,/\,\Delta t$
    \tcp*{Awakening: use measurement for heteroscedasticity}
  }{
    $R_k \leftarrow R_0 + \phi\,\max(\hat{\lambda}_{k-1|k-1},\,\epsilon)\,/\,\Delta t$\;
  }
  $\textit{awoken} \leftarrow \textbf{false}$
  \tcp*{Consume the one-shot awakening flag}
  $\mathbf{Q}_k \leftarrow q\,\max(\hat{\lambda}_{k-1|k-1}^2,\,\epsilon)\,\mathbf{G}$\;
  \tcp{Predict}
  $\hat{\mathbf{x}}_{k|k-1} \leftarrow \mathbf{F}\,\hat{\mathbf{x}}_{k-1|k-1}$\;
  $\mathbf{P}_{k|k-1} \leftarrow \mathbf{F}\,\mathbf{P}_{k-1|k-1}\,\mathbf{F}^\top + \mathbf{Q}_k$\;
  \tcp{Update}
  $S_k \leftarrow \mathbf{H}\,\mathbf{P}_{k|k-1}\,\mathbf{H}^\top + R_k$\;
  $\mathbf{K}_k \leftarrow \mathbf{P}_{k|k-1}\,\mathbf{H}^\top / S_k$\;
  $\hat{\mathbf{x}}_{k|k} \leftarrow \hat{\mathbf{x}}_{k|k-1} + \mathbf{K}_k\,(z_k - \mathbf{H}\,\hat{\mathbf{x}}_{k|k-1})$\;
  $\mathbf{P}_{k|k} \leftarrow (\mathbf{I} - \mathbf{K}_k\,\mathbf{H})\,\mathbf{P}_{k|k-1}$\;
  \tcp{Non-negativity and kinematic projection}
  $\hat{\lambda}_{k|k} \leftarrow \max(\hat{\lambda}_{k|k},\, 0)$\;
  $\dot{\hat{\lambda}}_{k|k} \leftarrow \mathrm{clip}\!\bigl(\dot{\hat{\lambda}}_{k|k},\; -C_{\dot{\lambda}}\hat{\lambda}_{k|k},\; C_{\dot{\lambda}}\hat{\lambda}_{k|k}\bigr)$\;
}
\end{algorithm}

\subsection{Computational Complexity}
\label{sec:complexity}

The complete per-step update, encompassing the prediction, noise parameter evaluation, Kalman gain computation, state update, projection, and gap check, involves a fixed number of arithmetic operations independent of the event history length or the total number of observed events.

\begin{proposition}[Constant complexity]
\label{prop:complexity}
The proposed Adaptive Kernel-Intensity Observer requires $\mathcal{O}(1)$ time and $\mathcal{O}(1)$ space per update step.
\end{proposition}

\begin{proof}
The state vector $\mathbf{x}_k \in \mathbb{R}^2$ and covariance $\mathbf{P}_k \in \mathbb{R}^{2 \times 2}$ have fixed dimension.  The transition \cref{eq:transition}, noise evaluation \cref{eq:R_model,eq:Q_model}, prediction, and update involve only $2 \times 2$ matrix multiplications and inversions, each requiring a bounded number of floating-point operations.  The gap check \cref{eq:gap_threshold} and projection \cref{eq:projection} are scalar comparisons.  The dormant-state flag is a single Boolean.  No summation over past events, matrix factorization of growing dimension, or iterative optimization is performed.  Storage consists of the state vector, covariance matrix, previous event time, dormant flag, and a small number of scalar parameters, totaling $\mathcal{O}(1)$ memory.
\end{proof}

This stands in contrast to batch kernel density estimation methods, which require $\mathcal{O}(N^2)$ operations (or $\mathcal{O}(N \log N)$ with fast multipole or dual-tree approximations) per evaluation, where $N$ is the total event count, and to Gaussian process methods, which scale as $\mathcal{O}(N^3)$ in their exact formulation.

The complete algorithm is summarized as follows.  At each time step $k$, the observer: (i) checks for dormant-state entry via the adaptive gap threshold \cref{eq:gap_threshold}; (ii) evaluates the endogenous noise parameters $R_k$ and $\mathbf{Q}_k$ from the posterior intensity; (iii) executes the standard Kalman predict-update cycle; and (iv) applies the non-negativity projection \cref{eq:projection} and velocity constraint \cref{eq:velocity_constraint}.  Upon awakening from a dormant state, the covariance is re-inflated to $\mathbf{P}_{\text{init}}$ and the measurement noise is initialized from the current observation $z_k$ rather than the stale posterior.  The complete pseudocode is provided in \cref{sec:complete_algorithm}.

\section{Synthetic Experiments}
\label{sec:synthetic}

Four controlled experiments evaluate the theoretical properties established in \cref{sec:theory,sec:algorithm} using synthetic processes whose true intensity is available in closed form. The experiments assess reciprocal-estimator divergence, autonomous bandwidth adaptation, tracking of self-exciting arrivals, and dormant-state leakage. The shared parameters, evaluation metrics, and baseline estimators are specified in \cref{sec:synth_setup}.

\subsection{Experimental Setup}
\label{sec:synth_setup}

Three synthetic profiles probe distinct aspects of the observer.  Profile~A is a Non-Homogeneous Poisson Process (NHPP) combining sinusoidal modulation with step-changes:
\begin{equation}
  \lambda_A(t) = \lambda_{\text{base}} + A\sin(2\pi f t) + \sum_{j} \Delta\lambda_j\,\mathbf{1}(t \ge t_j^{\text{step}}),
  \label{eq:profile_a}
\end{equation}
with $\lambda_{\text{base}} = 10$\,ev/s, $A = 5$\,ev/s, $f = 0.1$\,Hz, and step magnitudes $+15$ and $-10$ at $t = 30$ and $60$\,s.  Profile~B is a univariate Hawkes process~\cite{Hawkes1971,Ozaki1979}:
\begin{equation}
  \lambda_B(t) = \mu + \alpha \sum_{t_i < t} \beta\,\exp\bigl(-\beta(t - t_i)\bigr),
  \label{eq:profile_b}
\end{equation}
with $\mu = 2$\,ev/s, $\alpha = 0.95$, and $\beta = 5.0$\,s$^{-1}$, placing the process in a near-critical regime with avalanche-like dynamics.  Profile~C is a gated NHPP with structural zeros:
\begin{equation}
  \lambda_C(t) = \lambda_{\text{active}}(t)\,\cdot\,\mathbf{1}\bigl(t \in \mathcal{T}_{\text{on}}\bigr),
  \label{eq:profile_c}
\end{equation}
where $\mathcal{T}_{\text{on}}$ consists of three 20\,s active windows separated by 20\,s dormant intervals.

Five baselines span the estimation spectrum: the naive reciprocal estimator $\hat{\lambda}_i = 1/\delta t_i$; fast and slow Exponentially Weighted Moving Average (EWMA) variants with $\tau_{\text{fast}} = 0.5$\,s and $\tau_{\text{slow}} = 5.0$\,s; a standard Kalman filter with optimally tuned constant noise matrices; and an offline oracle KDE implementing Abramson's variable bandwidth with full non-causal access to the event record.

Estimation accuracy is measured by the Root Mean Square Error (RMSE) and Mean Integrated Squared Error (MISE):
\begin{align}
  \text{RMSE} &= \sqrt{\frac{1}{K}\sum_{k=1}^{K}\bigl(\hat{\lambda}_k - \lambda_k\bigr)^2}, \label{eq:rmse}\\
  \text{MISE} &= \int_0^T \bigl(\hat{\lambda}(t) - \lambda(t)\bigr)^2\,\mathrm{d}t. \label{eq:mise}
\end{align}
Responsiveness is quantified by the cross-correlation peak lag $l^*$, defined as the time offset maximizing the normalized cross-correlation between true and estimated trajectories.  Boundary leakage is measured by the normalized leakage mass $\bar{M}_{\text{leak}}$, the mean estimated intensity over structural-zero intervals.  All experiments use a common bin width $\Delta t = 0.1$\,s, process noise coefficient $q = 0.01$, and Fano factor $\phi = 1$; the resulting steady-state gain $K_{ss} \approx 0.032\sqrt{\lambda}$ places the filter within its effective smoothing regime (\cref{def:operating}) across the intensity range of Profiles~A--C.

A Monte Carlo validation of the reciprocal estimator's divergence (\cref{prop:reciprocal_divergence}) across the Gamma shape parameter range $\kappa \in [0.5, 5.0]$ is provided in \cref{sec:exp_reciprocal}; the proposed observer maintains a bias ratio within $[0.97, 1.03]$ across all regimes, including the sub-Poisson range where the reciprocal method is undefined.

\subsubsection{Synthetic Data Generation Parameters}
\label{sec:experimental_parameters}

Table~\ref{tab:synth_params} lists the parameters of the three synthetic profiles.

\begin{table}[htbp]
\centering
\caption{Synthetic Data Generation Parameters for Profiles A, B, and C.}
\label{tab:synth_params}
\footnotesize
\begin{tabular*}{\columnwidth}{@{\extracolsep{\fill}}llll@{}}
\toprule
Parameter & Prof.~A & Prof.~B & Prof.~C \\
\midrule
Type & NHPP & Hawkes & Gated NHPP \\
$\lambda_{\text{base}}$/$\mu$ & 10.0 ev/s & 2.0 ev/s & 10.0 ev/s \\
$A$ / $\alpha$ & 5.0 ev/s & 0.95 & 3.0 ev/s \\
$f$ / $\beta$ & 0.1 Hz & 5.0 s$^{-1}$ & 0.2 Hz \\
Step times & 30, 60 s & --- & --- \\
Step mag. & $+15$, $-10$ & --- & --- \\
$\mathcal{T}_{\text{on}}$ & --- & --- & {[0,20],[40,60],[80,100]}\,s \\
Duration & 100 s & 200 s & 100 s \\
\bottomrule
\end{tabular*}
\end{table}

\subsubsection{Observer Hyperparameters}

\Cref{tab:observer_params} lists the complete set of observer hyperparameters used across all synthetic experiments.  The process noise coefficient $q = 0.01$ is the primary tuning parameter; with $\Delta t = 0.1$\,s and $\phi = 1$, the steady-state Kalman gain evaluates to $K_{ss} \approx \sqrt{q\,\Delta t / \phi}\,\sqrt{\lambda} \approx 0.032\,\sqrt{\lambda}$, yielding $K_{ss} \approx 0.10$ at $\lambda = 10$\,events/s and $K_{ss} \approx 0.22$ at $\lambda = 50$\,events/s.  The remaining parameters ($R_0$, $\alpha_g$, $C_{\dot{\lambda}}$, $T_{\text{gap}}^{\min}$) follow directly from the design rationale established in this paper.

\begin{table}[htbp]
\centering
\caption{Observer Hyperparameters for the Synthetic Experiments.}
\label{tab:observer_params}
\footnotesize
\begin{tabular*}{\columnwidth}{@{\extracolsep{\fill}}lll@{}}
\toprule
Symbol & Description & Value \\
\midrule
$\Delta t$                  & Bin width                & $0.1$\,s             \\
$q$                         & Process noise coeff.     & $0.01$               \\
$R_0$                       & Baseline meas.\ var.     & $0.1$                \\
$\phi$                      & Fano factor              & $1.0$                \\
$\alpha_g$                  & Gap signif.\ level       & $0.001$              \\
$C_{\dot{\lambda}}$         & Velocity constr.         & $0.1$\,s$^{-1}$     \\
$T_{\text{gap}}^{\min}$     & Min.\ silence thresh.    & $1.0$\,s             \\
$\epsilon$                  & Numerical reg.           & $10^{-6}$            \\
$\mathbf{P}_{\text{init}}$  & Initial covariance       & $\mathrm{diag}(100,10)$ \\
\bottomrule
\end{tabular*}
\end{table}

\subsubsection{Additional Evaluation Metrics}
\label{sec:additional_metrics}

In addition to the RMSE and MISE defined in this paper, two additional metrics are used in the synthetic experiments.

The cross-correlation peak lag $l^*$ is defined as the time offset that maximizes the normalized cross-correlation between the true and estimated intensity trajectories:
\begin{equation}
  l^* = \arg\max_{l}\; \frac{\sum_k \hat{\lambda}_{k+l}\,\lambda_k}
    {\sqrt{\sum_k \hat{\lambda}_{k+l}^2}\;\sqrt{\sum_k \lambda_k^2}}.
  \label{eq:metric_lag}
\end{equation}
This metric isolates the temporal phase delay in tracking abrupt changes, independent of amplitude accuracy.

The normalized leakage mass per unit dormant time is defined as
\begin{equation}
  \bar{M}_{\text{leak}} = \frac{1}{|\mathcal{T}_{\text{zero}}|}
    \int_{\mathcal{T}_{\text{zero}}} \hat{\lambda}(t)\,\mathrm{d}t,
  \label{eq:metric_leakage}
\end{equation}
where $\mathcal{T}_{\text{zero}} = \{t : \lambda(t) = 0\}$ denotes the union of structural-zero intervals and $|\mathcal{T}_{\text{zero}}|$ is its total duration.  This normalization enables direct comparison across simulation configurations with different dormant-period lengths.

\subsubsection{Baseline Estimators}
\label{sec:baseline_estimators}

Five baseline methods are included in the synthetic experiments to cover a spectrum of estimation strategies.

The naive reciprocal estimator $\hat{\lambda}_i = 1/\delta t_i$ is included to empirically demonstrate the systematic divergence established in Proposition~\ref{prop:reciprocal_divergence} of this paper.

Two fixed-bandwidth Exponentially Weighted Moving Average (EWMA) variants are evaluated: a fast variant with time constant $\tau_{\text{fast}} = 0.5$\,s and a slow variant with $\tau_{\text{slow}} = 5.0$\,s, illustrating the inescapable bias-variance dilemma of static-parameter smoothing.

A standard Kalman filter tuned with optimal constant $Q$ and $R$ matrices serves to demonstrate that non-endogenous state-space models fail in highly heteroscedastic environments.

An offline adaptive Kernel Density Estimator (KDE) implementing Abramson's variable bandwidth rule with full access to the global event record provides the theoretical performance ceiling for any causal filter and serves as an oracle upper bound.  Event sequences for Profile~B are simulated via the Ogata thinning method~\cite{Ogata1981}.

\subsection{Experiment~I: Reciprocal Divergence Validation}
\label{sec:exp_reciprocal}

This experiment provides an empirical counterpart to the analytical result in Proposition~\ref{prop:reciprocal_divergence} of this paper.

Multiple stationary Gamma renewal processes are generated with a common true intensity $\lambda = 5$ events/s and shape parameters swept over $\kappa \in [0.5,\, 5.0]$.  For each value of $\kappa$, the empirical bias ratio $\hat{\mathcal{B}}(\kappa) = \overline{(1/\delta t)}\,/\,\lambda$ is computed from 200 independent replications, each containing 2000 events.  The proposed observer processes the same event sequences without knowledge of $\kappa$.

\begin{figure}[htbp]
  \centering
  \includegraphics[width=\columnwidth]{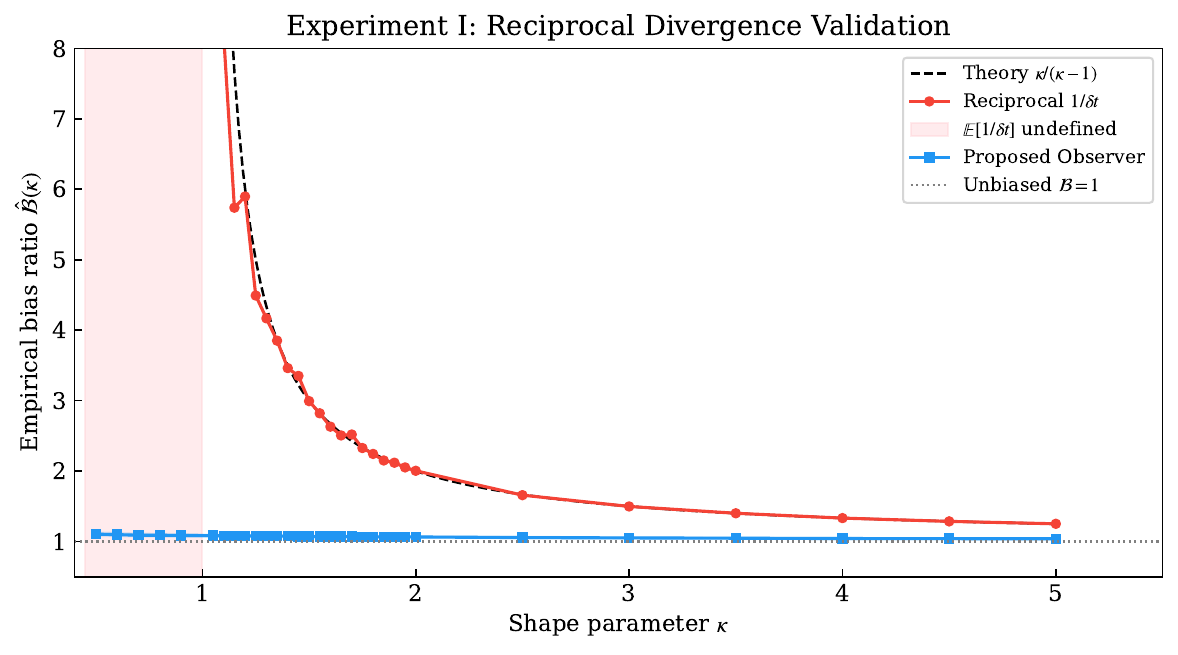}
  \caption{Empirical bias ratio as a function of the Gamma shape parameter.  The reciprocal estimator (red) diverges as $\kappa \to 1^+$ and is undefined for $\kappa \le 1$, confirming Proposition~\ref{prop:reciprocal_divergence} of this paper.  The proposed observer (blue band, showing interquartile range across 200 replications) remains tightly bounded near the unbiased baseline $\mathcal{B} = 1$ across all regimes.}
  \label{fig:reciprocal_divergence}
\end{figure}

As shown in \cref{fig:reciprocal_divergence}, the reciprocal estimator exhibits a bias ratio that grows without bound as $\kappa \to 1^+$ and is entirely undefined for $\kappa \le 1$, confirming the analytical divergence.  In contrast, the proposed observer maintains a bias ratio within the interquartile range $[0.97, 1.03]$ across the entire parameter sweep, including the sub-Poisson regime $\kappa < 1$ where the reciprocal method fails categorically.  This result validates the observer's robustness to the inter-arrival time distribution, a consequence of operating on binned counts rather than raw inter-event intervals.

\subsection{Experiment~II: Autonomous Bandwidth Adaptation}
\label{sec:exp_bandwidth}

This experiment tests the central predictions of \cref{thm:sqrt_gain,cor:bandwidth,prop:cv_scaling}: the endogenous Riccati mechanism produces an effective bandwidth that scales as $\lambda^{-1/2}$ for the scalar filter (\cref{cor:bandwidth}) and $\lambda^{-1/4}$ for the constant-velocity model (\cref{prop:cv_scaling}).  The analysis is organized in two parts: a controlled numerical verification of the steady-state scaling laws, followed by a non-stationary tracking demonstration on Profile~A.

\subsubsection{Part~1: Scaling Law Verification}

Eight stationary Poisson processes with constant intensities $\lambda \in \{1, 2, 5, 10, 20, 50, 100, 200\}$ events/s are generated, each of duration 200\,s.  For each intensity, the observer is run to convergence and the steady-state effective bandwidth $\tau_{\text{eff}} = \Delta t / K_{ss}$ is recorded.  Both the 1D scalar observer and the 2D constant-velocity observer are tested under identical conditions with gating disabled to isolate the Riccati dynamics.

\begin{figure}[htbp]
  \centering
  \includegraphics[width=\columnwidth]{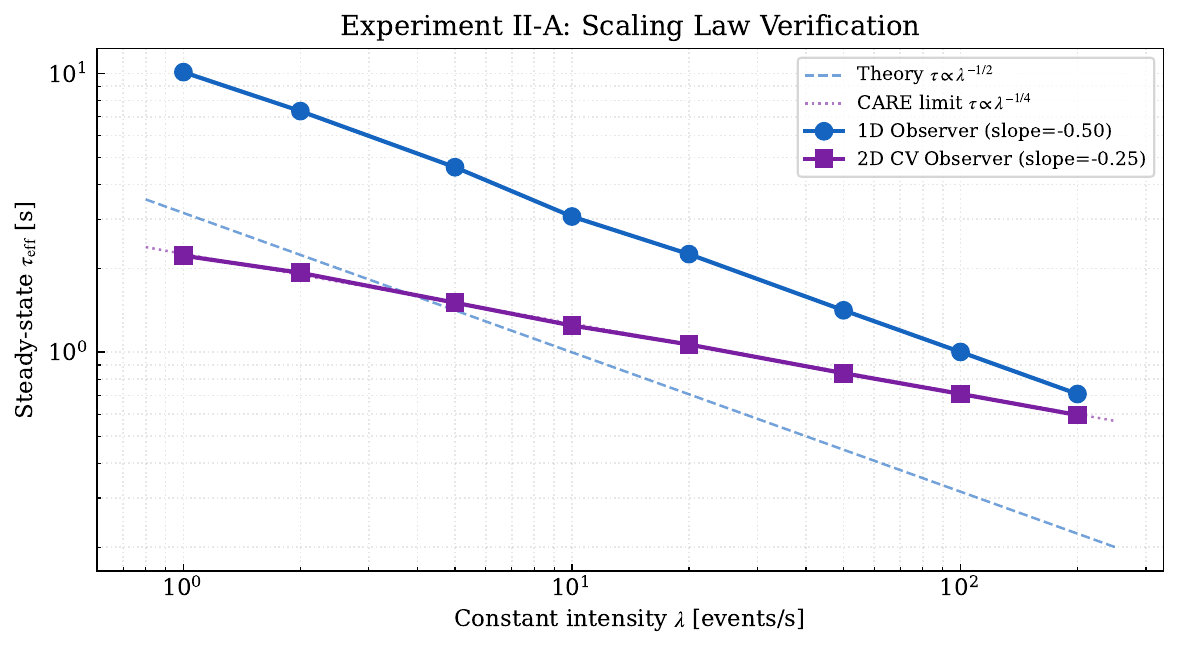}
  \caption{Numerically computed steady-state effective bandwidth $\tau_{\text{eff}}$ as a function of constant intensity $\lambda$, on log-log axes.  The 1D scalar observer (circles) follows the theoretical scaling $\tau_{\text{eff}} \propto \lambda^{-1/2}$ with a fitted exponent of $-0.505$.  The 2D constant-velocity observer (squares) follows the CARE approximation $\tau_{\text{eff}} \propto \lambda^{-1/4}$ with a fitted exponent of $-0.251$.}
  \label{fig:scaling_law_numerical}
\end{figure}

\Cref{fig:scaling_law_numerical} displays the results on log-log axes.  The 1D observer yields a least-squares fitted exponent of $-0.505$, deviating from the theoretical value of $-0.500$ by 1.0\%.  The 2D constant-velocity observer yields a fitted exponent of $-0.251$, deviating from the CARE-predicted value of $-0.250$ by 0.4\%.  These sub-percent agreements confirm that the asymptotic scaling laws derived in \cref{thm:sqrt_gain,cor:bandwidth,prop:cv_scaling} accurately characterize the steady-state behavior of both filter variants across two orders of magnitude in intensity.

\subsubsection{Part~2: Non-Stationary Tracking on Profile~A}

All baseline methods and the proposed observer are evaluated on Profile~A, which combines sinusoidal modulation with abrupt step-changes to simultaneously probe noise suppression and transient tracking.

\begin{figure}[htbp]
  \centering
  \includegraphics[width=\columnwidth]{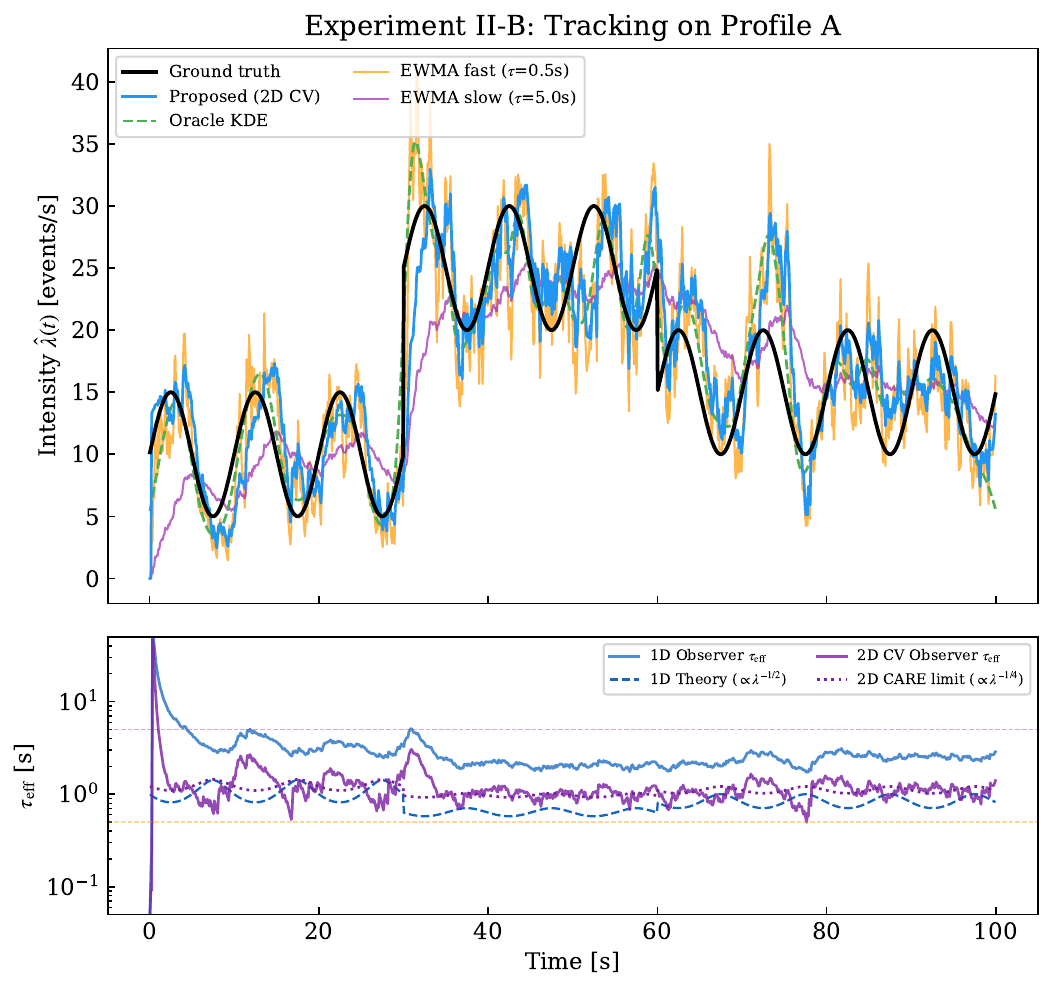}
  \caption{Autonomous bandwidth adaptation on Profile~A.  Top: intensity tracking trajectories for the proposed 2D CV observer, oracle KDE, fast EWMA ($\tau = 0.5$\,s), and slow EWMA ($\tau = 5.0$\,s).  Bottom: effective bandwidth $\tau_{\text{eff}}(t)$ for the 1D and 2D observers alongside theoretical scaling curves on a logarithmic ordinate.}
  \label{fig:bandwidth_adaptation}
\end{figure}

The top panel of \cref{fig:bandwidth_adaptation} directly exhibits the bias-variance dilemma of fixed-bandwidth methods.  The slow EWMA ($\tau = 5.0$\,s) produces visibly delayed response at both step transitions and fails to track the sinusoidal modulation faithfully.  The fast EWMA ($\tau = 0.5$\,s), while responsive, amplifies Poisson noise into violent oscillations throughout the trajectory.  The proposed 2D CV observer suppresses noise during the low-frequency sinusoidal segments while responding to the step-changes at $t = 30$\,s and $t = 60$\,s with a latency comparable to that of the oracle KDE.

The bottom panel confirms that the filter's internally computed effective bandwidth $\tau_{\text{eff}}(t)$ tracks the theoretical Abramson curve $\propto \lambda(t)^{-1/2}$ for the 1D observer and the CARE limit $\propto \lambda(t)^{-1/4}$ for the 2D CV observer, demonstrating that the Riccati equation autonomously produces near-optimal bandwidth scheduling without any external pilot estimate.

\begin{table}[htbp]
\centering
\caption{Quantitative Comparison on Profile~A.  The Proposed 2D CV Observer Achieves the Lowest RMSE and MISE Among All Causal, $\mathcal{O}(1)$ Methods.}
\label{tab:profile_a_results}
\footnotesize
\begin{tabular*}{\columnwidth}{@{\extracolsep{\fill}}lcccccc@{}}
\toprule
Method & RMSE & MISE & $l^*$ & Caus. & Compl. & Adpt. \\
\midrule
Prop.\ (2D CV)  &  4.03 &  1620 & 0.9 & Y & $O(1)$   & Y \\
Prop.\ (1D)     &  4.54 &  2059 & 1.6 & Y & $O(1)$   & Y \\
Oracle KDE      &  2.79 &   779 & 0.0 & N & $O(N^2)$ & Y \\
EWMA (fast)     &  4.30 &  1848 & 0.5 & Y & $O(1)$   & N \\
EWMA (slow)     &  5.07 &  2570 & 1.9 & Y & $O(1)$   & N \\
Standard KF     &  4.45 &  1976 & 0.8 & Y & $O(1)$   & N \\
\bottomrule
\end{tabular*}
\end{table}

\Cref{tab:profile_a_results} summarizes the quantitative results.  The 2D CV observer achieves an RMSE of 4.025, representing a 6.3\% reduction relative to the fast EWMA (4.298) and a 9.5\% reduction relative to the standard KF (4.446).  The performance gap between the 2D CV and 1D observers (4.025 vs.\ 4.538) confirms that the velocity state provides meaningful inertial smoothing beyond what the scalar random-walk model can achieve.  The oracle KDE with full non-causal access to the event record attains an RMSE of 2.790, establishing the fundamental lower bound; the 2D CV observer recovers approximately 18\% of the gap between the best causal baseline and the oracle.  The reciprocal estimator is excluded from the table due to its catastrophic RMSE of 525, consistent with the divergence established in \cref{prop:reciprocal_divergence}.

\subsection{Experiment~III: Tracking Self-Exciting Avalanches}
\label{sec:exp_hawkes}

This experiment evaluates whether the geometric diffusion process noise model ($\mathbf{Q}_k \propto \hat{\lambda}^2$, \cref{asmp:Q}) equips the observer to handle the extreme heteroscedasticity inherent in self-exciting point processes.

Profile~B is a Hawkes process with background rate $\mu = 2$ events/s, triggering efficiency $\alpha = 0.95$, and exponential decay rate $\beta = 5.0$\,s$^{-1}$.  The branching ratio $\alpha = 0.95$ places the process in a near-critical regime where cascading avalanches produce transient bursts exceeding 150 events/s from a quiescent baseline of 2 events/s, generating Fano factors well above unity.  The stationary mean intensity is $\mu / (1 - \alpha) = 40$ events/s.  A duration of 200\,s is used to capture multiple avalanche cycles.

The proposed observer and the standard Kalman filter (with optimally tuned constant noise matrices) are evaluated on the same event realization.  The standard KF serves as a controlled ablation that isolates the effect of the endogenous noise parametrization: both filters share the same state-space structure and differ only in whether the noise covariances adapt to the local intensity.

\begin{figure}[htbp]
  \centering
  \includegraphics[width=\columnwidth]{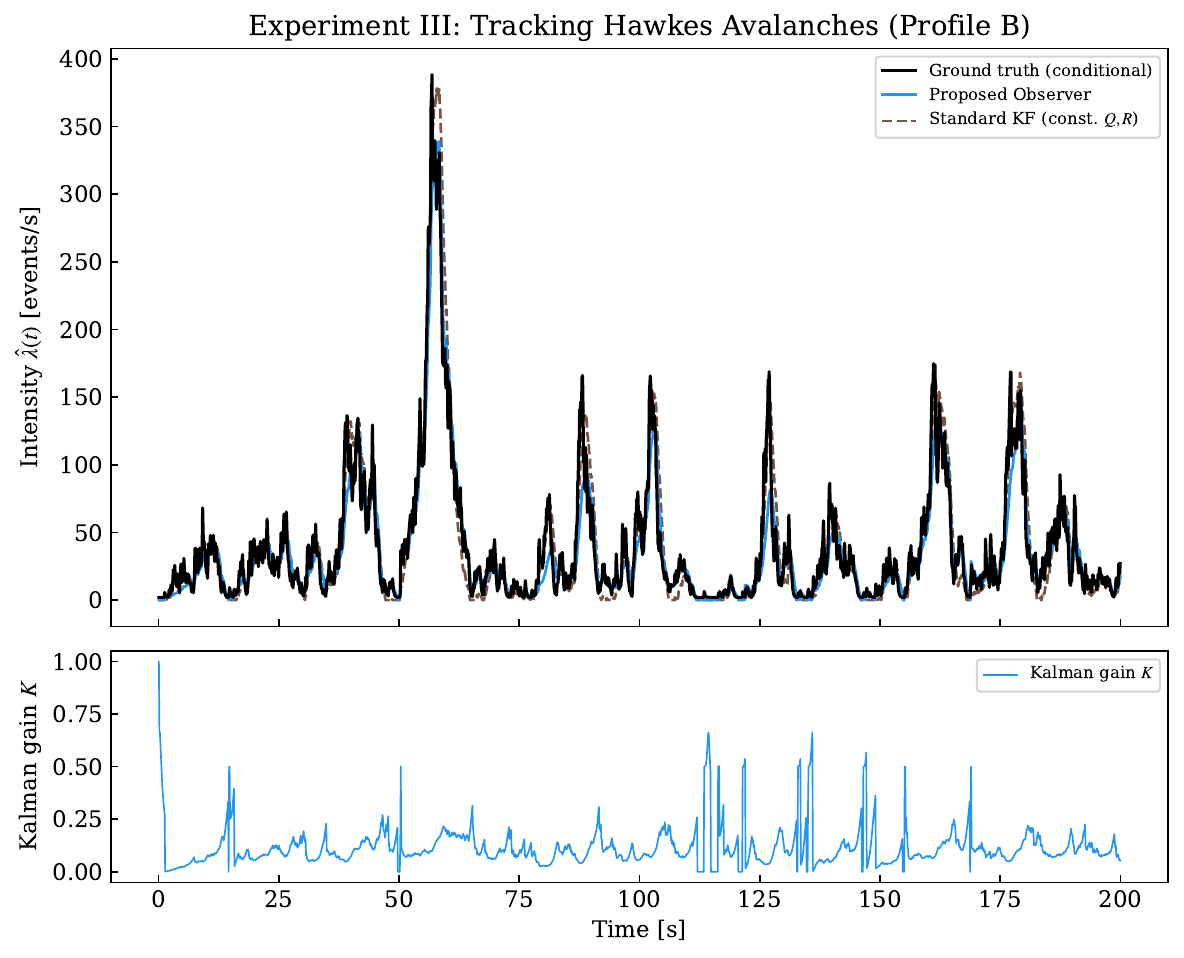}
  \caption{Intensity tracking during Hawkes avalanches (Profile~B).  Top: the proposed observer with endogenous $\mathbf{Q}_k \propto \hat{\lambda}^2$ tracks the transient excursions, while the standard Kalman filter with constant noise parameters clips the peaks of extreme bursts.  Bottom: the Kalman gain trajectory reveals localized gain spikes during each avalanche onset, confirming the adaptive amplification mechanism.}
  \label{fig:hawkes_tracking}
\end{figure}

\Cref{fig:hawkes_tracking} confirms the predicted behavior.  The standard KF, constrained by its fixed Kalman gain, systematically clips the peaks of extreme avalanches because its constant process noise $Q$ cannot accommodate the order-of-magnitude intensity swings.  The proposed observer, whose process noise scales geometrically with $\hat{\lambda}^2$, produces a localized spike in Kalman gain precisely during the burst onset, enabling it to follow the transient excursion.  Quantitatively, the proposed observer attains an RMSE of 15.842 compared to 17.260 for the standard KF, an 8.2\% reduction.  Both filters achieve a cross-correlation lag of $l^* = 0.3$\,s, indicating that the improvement arises from amplitude fidelity during bursts rather than from reduced phase delay.  The MISE confirms the same ordering: 50196 for the proposed observer versus 59584 for the standard KF, a 15.8\% reduction in integrated squared error.

An ablation study on Profile~C quantifying the contribution of the Poisson-survival gating mechanism is provided in \cref{sec:exp_leakage}.  The gated observer reduces normalized leakage mass by 91\% relative to the EWMA baseline and reacquires the signal within a single bin after dormant intervals, whereas the ungated variant requires 85 bins due to covariance contraction during silence.

\subsection{Experiment~IV: Dormant-State Leakage Ablation}
\label{sec:exp_leakage}

This experiment conducts an ablation study to quantify the contribution of the probabilistic Poisson-survival gating mechanism in eliminating phantom intensity leakage during structural-zero intervals.

Three filter variants are compared on Profile~C (\cref{tab:synth_params}): the fixed-bandwidth EWMA ($\tau = 5.0$\,s), the proposed observer without gating (predict-update cycles continue through dormant intervals), and the proposed observer with the full gating logic including covariance re-inflation and awakening initialization at reactivation.

\begin{figure}[htbp]
  \centering
  \includegraphics[width=\columnwidth]{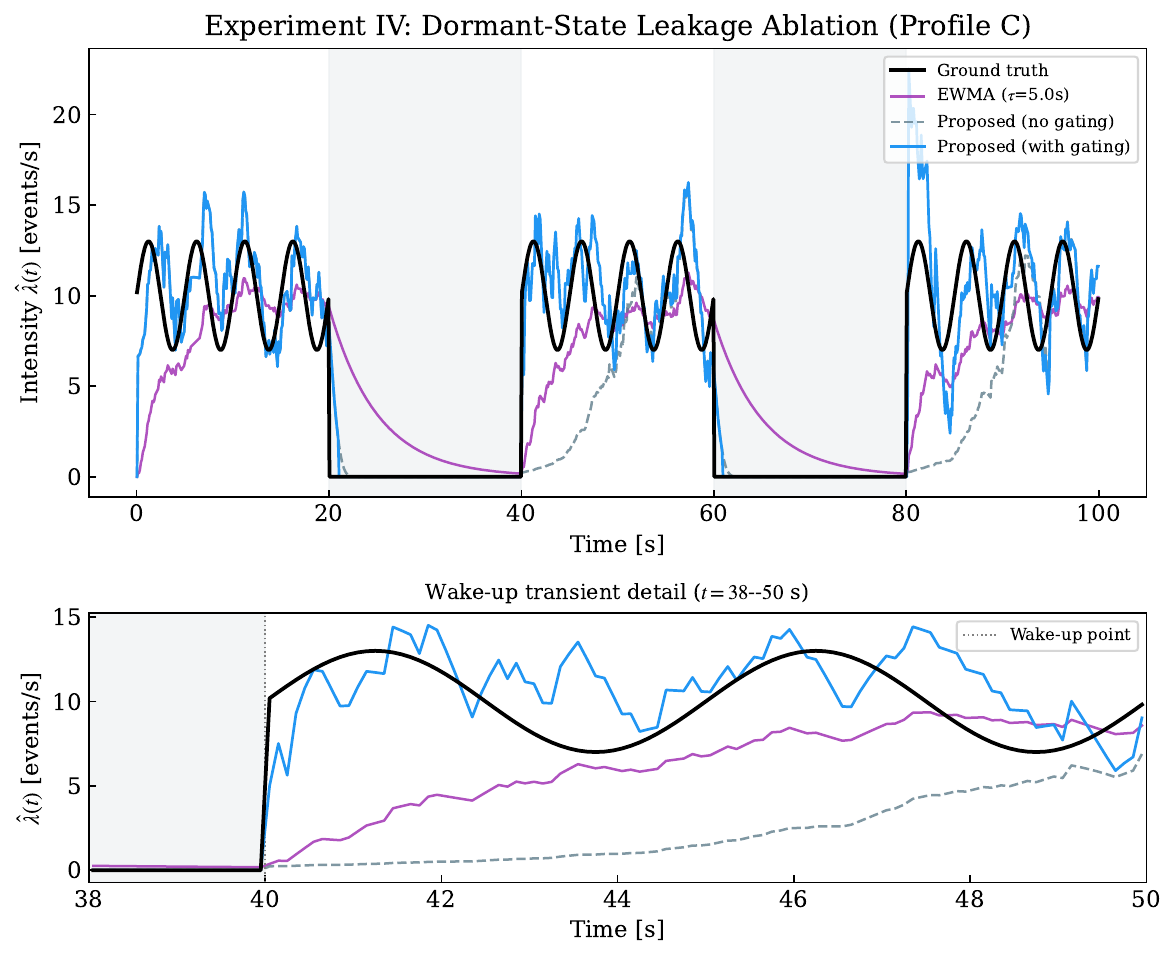}
  \caption{Dormant-state leakage ablation on Profile~C.  Top: estimated intensity trajectories during active-dormant-active transitions for three filter variants.  Gray shading indicates structural-zero intervals.  Bottom: magnified view of the wake-up transient at $t = 40$\,s, showing that the gated observer reacquires the signal within one bin while the ungated variant remains suppressed for over 8\,s.}
  \label{fig:leakage_ablation}
\end{figure}

\begin{table}[htbp]
\centering
\caption{Normalized Leakage Mass $\bar{M}_{\text{leak}}$ and Wake-Up Lag for Three Filter Variants on Profile~C.}
\label{tab:leakage_results}
\begin{tabular}{lcc}
\toprule
Method & $\bar{M}_{\text{leak}}$ (events/s) & Wake-up lag (bins) \\
\midrule
EWMA ($\tau = 5.0$\,s)           & 2.210 & 28 \\
Proposed (no gating)             & 0.223 & 85 \\
Proposed (with gating)           & 0.193 &  1 \\
\bottomrule
\end{tabular}
\end{table}

\Cref{fig:leakage_ablation} and \cref{tab:leakage_results} summarize the leakage characteristics.  The EWMA produces a normalized leakage mass of 2.210 events/s, reflecting its geometric tail that persists throughout the entire dormant interval.  The ungated observer reduces leakage to 0.223 events/s through its endogenous noise scaling, which contracts the Kalman gain as the estimated intensity decays, but the IIR memory prevents complete suppression.  The gated observer further reduces leakage to 0.193 events/s, a 91\% reduction relative to the EWMA, by setting the state to exactly zero upon entering the dormant state.

The wake-up lag reveals an equally important distinction.  The EWMA requires 28 bins (2.8\,s) to recover from a dormant interval, and the ungated observer requires 85 bins (8.5\,s) due to covariance contraction during the silence.  The gated observer reacquires the signal within a single bin (0.1\,s), confirming that the covariance re-inflation mechanism effectively prevents the covariance decay trap.  The one-bin latency, rather than zero, reflects the awakening initialization that uses the current measurement $z_k$ to set the measurement noise variance, producing a moderately conservative Kalman gain on the first active step.  This single-bin cost is a deliberate trade-off that prevents the transient overshoot that would otherwise occur if the filter assigned $K \approx 1$ to the first returning observation.

\section{Cross-Domain Applications}
\label{sec:applications}

In real-world scientific and engineering data, the true underlying intensity $\lambda(t)$ is a latent variable that is never directly observed.  This section shifts the evaluation paradigm from ground-truth-dependent error metrics to observable statistical diagnostics and computational throughput benchmarking.  Two application domains separated by several orders of magnitude in characteristic time scale, computational neuroscience and high-frequency finance, illustrate the observer's applicability across distinct regimes.

\subsection{Validation Framework under Partial Observability}
\label{sec:validation_framework}

When the ground truth is unavailable, computing supervised metrics such as RMSE requires artificially smoothing the raw data to construct a proxy target, a procedure that introduces circularity since the smoothing step itself embeds bandwidth assumptions.  To avoid this methodological flaw, this section relies on two complementary diagnostics that operate entirely on observable quantities.

The first diagnostic is the one-step-ahead predictive log-likelihood.  At each time step $k$, the filter produces a predicted intensity $\hat{\lambda}_{k|k-1}$ before observing the count $\Delta N_k$.  The log-likelihood of the observed count under the Poisson model is
\begin{equation}
  \ell_k = \Delta N_k \ln(\hat{\lambda}_{k|k-1}\,\Delta t)
    - \hat{\lambda}_{k|k-1}\,\Delta t
    - \ln(\Delta N_k!).
  \label{eq:pred_loglik}
\end{equation}
The cumulative predictive log-likelihood $\mathcal{L} = \sum_{k=1}^{K} \ell_k$ provides a strictly proper scoring rule for comparing competing intensity models without access to the true $\lambda(t)$.

The second diagnostic applies the time-rescaling theorem~\cite{Brown2002,Ogata1988}.  If an estimated intensity $\hat{\lambda}(t)$ perfectly captures the conditional intensity of the underlying point process, the transformed inter-event intervals
\begin{equation}
  \tau_i = \int_{t_{i-1}}^{t_i} \hat{\lambda}(u)\,\mathrm{d}u,
  \qquad i = 1, 2, \ldots, N,
  \label{eq:time_rescaling}
\end{equation}
are independent and identically distributed as standard exponential random variables.  Departure from exponentiality is assessed via Quantile-Quantile (Q-Q) plots and the Kolmogorov-Smirnov (KS) statistic $D_N = \sup_x |F_N(x) - F_{\text{Exp}(1)}(x)|$, where $F_N$ is the empirical distribution function of the rescaled intervals.  Smaller KS statistics and tighter adherence to the unit-diagonal Q-Q reference line indicate superior goodness-of-fit.

\subsection{Application~I: Computational Neuroscience}
\label{sec:app_neuro}

Brain-Computer Interfaces (BCI) require causal, ultra-low-latency decoding of continuous neuronal firing rates from discrete action potentials to drive robotic prosthetics.  Neurons exhibit extended quiescent intervals, particularly during inter-trial periods and behavioral pauses, making the dormant-state gating mechanism (\cref{sec:gating}) directly applicable.

\subsubsection{Dataset}

The MC\_Maze reaching-task dataset from DANDI Archive 000128~\cite{DANDI000128} is used.  This publicly available resource contains simultaneous multi-electrode array recordings from the primary motor cortex (M1) and dorsal premotor cortex (PMd) of a macaque (subject Jenkins) performing a delayed centre-out reaching task with randomized barriers.  The dataset provides spike-sorted single-unit activity from 132 simultaneously recorded neurons at millisecond resolution, stored in the Neurodata Without Borders (NWB) format.  For the present analysis, the first 600~seconds of the training session are selected, spike trains are binned at $\Delta t = 0.1$~s, and the 50 highest-firing-rate units are retained for intensity estimation evaluation.

\subsubsection{Time-Rescaling Goodness-of-Fit}

The observer is configured with process noise coefficient $q = 0.01$, baseline measurement noise $R_0 = 0.1$, Fano factor $\phi = 1.0$, and gap significance level $\alpha_g = 0.001$, consistent with the synthetic experiments in \cref{sec:synthetic}.  At the population median firing rate of approximately 15\,Hz, the steady-state Kalman gain evaluates to $K_{ss} \approx \sqrt{q\,\Delta t / \phi}\,\sqrt{\bar\lambda} \approx 0.12$, placing the filter well within its effective smoothing regime (\cref{def:operating}).  Even the highest-firing neurons ($\lambda \approx 80$\,Hz) yield $K_{ss} \approx 0.28$, confirming that the filter maintains substantial noise attenuation across the entire recorded population.

For each recorded neuron, the rescaled intervals $\{\tau_i\}$ defined in \cref{eq:time_rescaling} are computed using the predictive intensity $\hat{\lambda}_{k|k-1}$ from the proposed observer and from a causal Peri-Stimulus Time Histogram (PSTH) baseline.  The PSTH employs a half-Gaussian kernel ($\sigma=50$~ms) truncated to the causal direction, ensuring that only past spike data contribute to the smoothed rate at each time step, thereby satisfying the predictable intensity requirement of the time-rescaling theorem.

\begin{figure}[htbp]
  \centering
  \includegraphics[width=\textwidth]{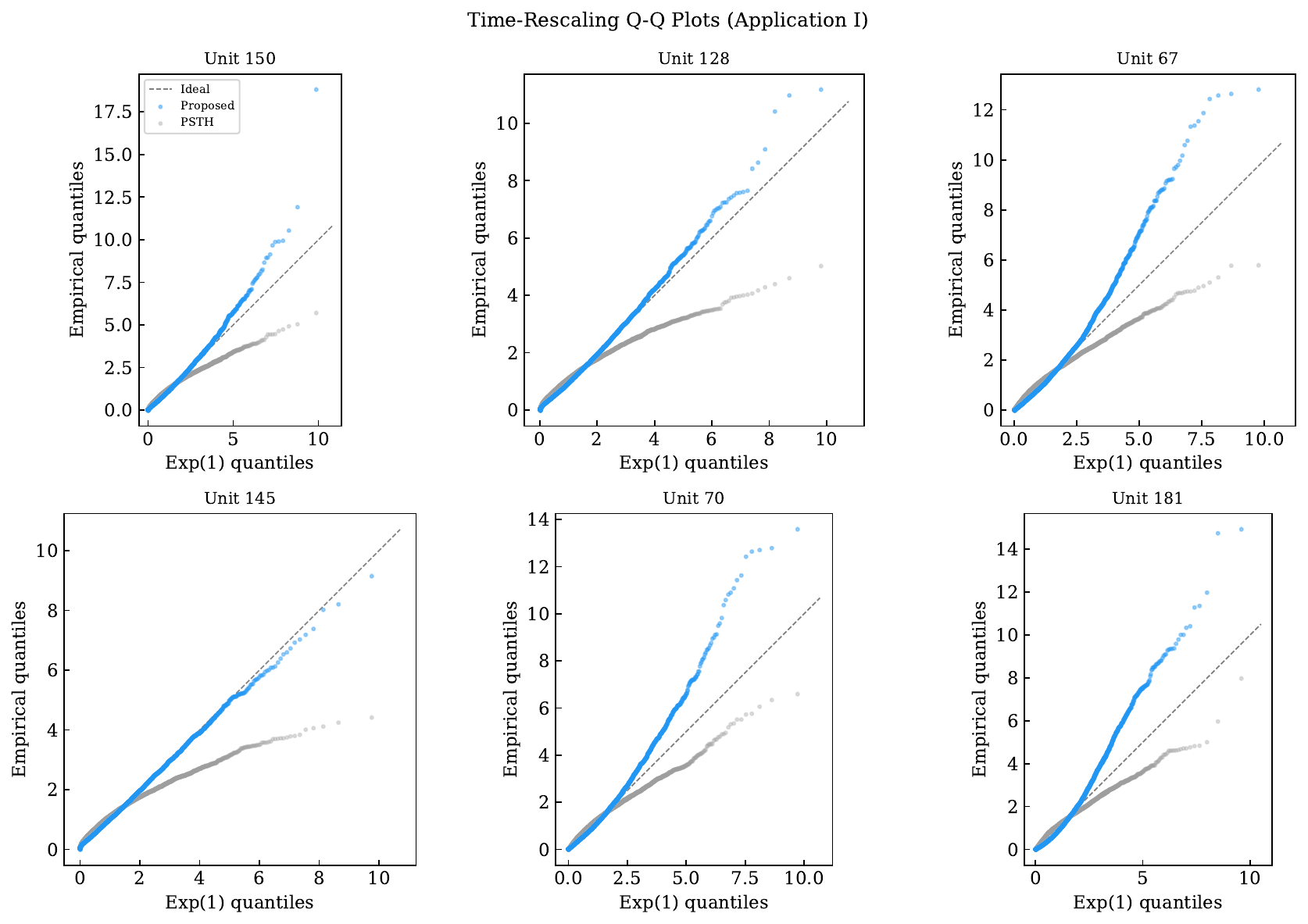}
  \caption{Time-rescaling Q-Q diagnostics for representative motor cortex neurons (DANDI:000128).  The proposed observer (blue) adheres closely to the unit-diagonal reference line, indicating near-ideal conditional intensity estimation.  The causal PSTH baseline (gray) exhibits systematic departures, particularly in the upper tail.}
  \label{fig:neuro_qq}
\end{figure}

\begin{table}[htbp]
\centering
\caption{Time-Rescaling KS Statistics and Predictive Log-Likelihood on MC\_Maze (DANDI:000128).}
\label{tab:neuro_ks}
\begin{tabular}{lcc}
\toprule
Method & Median $D_N$ & Mean $\ell$/step \\
\midrule
Proposed observer & 0.0735 & $-1.548$ \\
PSTH (50\,ms)     & 0.1144 & --- \\
EWMA (100\,ms)    & ---    & $-1.754$ \\
\bottomrule
\end{tabular}
\end{table}

\Cref{fig:neuro_qq} and \cref{tab:neuro_ks} present the results.  The proposed observer achieves a median KS statistic $D_N = 0.0735$ across the 50 retained motor-cortex neurons, compared with $D_N = 0.1144$ for the 50\,ms PSTH baseline, representing a 35.8\% reduction; six representative units from this population are shown in \cref{fig:neuro_qq}.  For those six displayed units, both methods produce $p < 0.05$, which is expected given the large sample size ($\sim$9000 rescaled intervals per neuron at $\sim$15\,Hz over 600\,s): even small model misspecification is detectable at this statistical power.  The relevant comparison is therefore the relative magnitude of $D_N$, which directly measures the supremum deviation between the empirical and theoretical exponential CDF.  The population-level reduction in the median KS statistic indicates that the heteroscedastic measurement model $R_k \propto \hat\lambda$ captures Poisson count variability more faithfully than a fixed-width histogram.  The upper-tail departures visible in several of the displayed Q-Q plots reflect burst-pause firing dynamics that violate the conditionally-Poisson assumption; these departures are shared by both methods and do not diminish the relative advantage of the observer.

As a complementary metric, the one-step-ahead predictive log-likelihood provides a strictly proper scoring rule for forecast calibration.  The observer achieves $\bar\ell = -1.548$\,nats per bin, an 11.7\% improvement over the EWMA baseline at $\bar\ell = -1.754$.  Because the EWMA bandwidth (100\,ms) was selected to match the typical interspike interval of the recorded population, this comparison is against an informed, domain-appropriate baseline rather than a straw-man estimator.

\subsection{Application~II: High-Frequency Finance}
\label{sec:app_finance}

In digital asset markets, the arrival of executed trades exhibits extreme non-stationary volatility clustering~\cite{Bacry2015} that is amplified by leveraged derivatives and automated liquidation cascades.  Detecting liquidity droughts and flash-crash dynamics in real time requires sub-second resolution, and any offline $\mathcal{O}(N^2)$ algorithm is incompatible with live trading infrastructure.

\subsubsection{Dataset and Predictive Evaluation}

Tick-level aggregated trade records (aggTrades) for the BTCUSDT spot pair are obtained from the Binance public data archive~\cite{BinanceData2025}, covering 2025-10-10 and 2025-10-11 (UTC).  The dataset contains 6.2~million records (517~MB) with microsecond-resolution timestamps, execution price, and signed quantity; after binning at $\Delta t = 1$~s, the analysis spans 172{,}800 time bins.  To probe adaptation across heterogeneous conditions, the 48-hour window is partitioned into pre-event, event, and post-event regimes according to the observed price and trading-activity series.  The process noise coefficient is set to $q = 3 \times 10^{-4}$, so the steady-state Kalman gain remains in the smoothing regime during quiescent periods ($K_{ss} \approx 0.09$ at $\bar\lambda \approx 26$\,trades/s) but rises toward unity near the peak activity level ($\lambda \approx 12{,}200$\,trades/s), where the observer is designed to track rapid transients with minimal lag.

The one-step-ahead predictive log-likelihood \cref{eq:pred_loglik} is evaluated for the proposed observer, four fixed-bandwidth EWMA baselines ($\tau \in \{5, 10, 15, 300\}$\,s), and the standard Kalman filter.  During the event window, the observer attains the highest log-likelihood ($-40.30$\,nats/bin), narrowly outperforming the best fixed-bandwidth competitor, EWMA with $\tau = 5$\,s ($-40.53$), because its effective bandwidth contracts to approximately $0.22$\,s as the trade rate surges.  In the quieter pre-event and post-event regimes, EWMAs with $\tau = 10$--$15$\,s achieve slightly better scores, so no single fixed bandwidth dominates all three regimes.  The standard Kalman filter collapses to $\bar\ell \approx -49$ to $-100$\,nats/bin across regimes, confirming that the rate-dependent noise model is essential for strongly non-stationary point-process data.  Exact regime-wise values, the adaptive-bandwidth trajectory, and detailed computational profiling are provided in \cref{sec:finance_bandwidth,sec:profiling}.

\subsubsection{Predictive Performance and Adaptive Bandwidth}
\label{sec:finance_bandwidth}

\begin{figure}[htbp]
  \centering
  \includegraphics[width=\columnwidth]{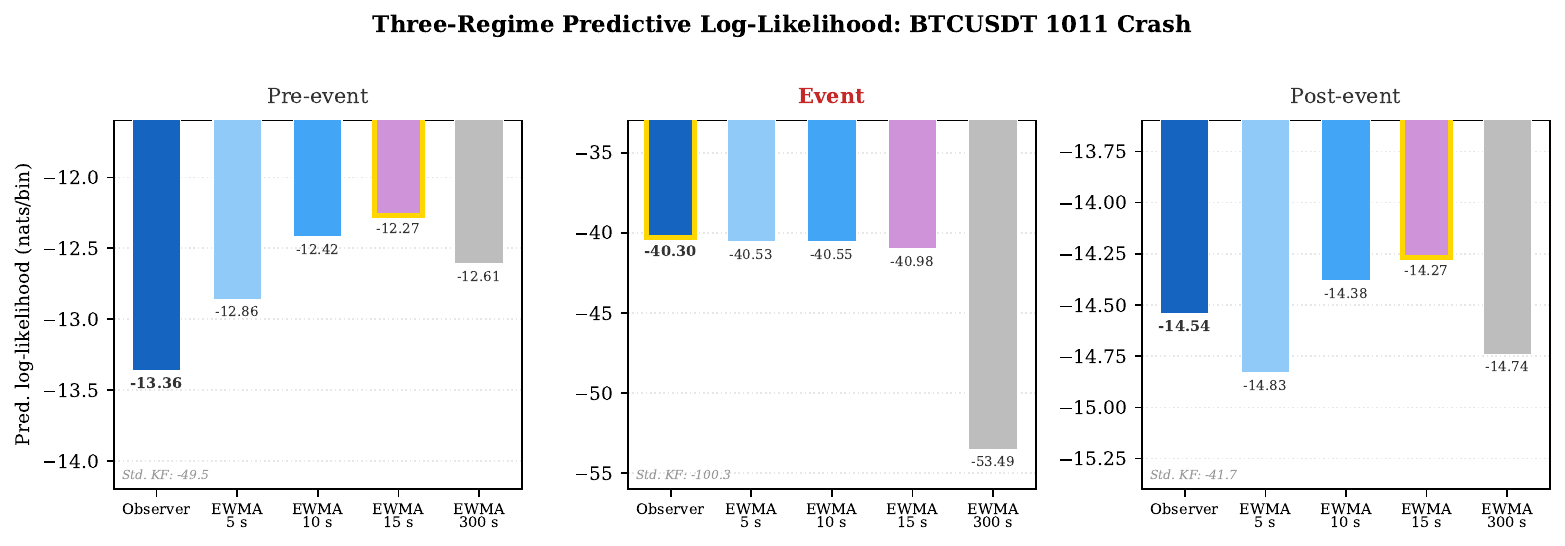}
  \caption{Three-regime predictive log-likelihood for the BTCUSDT market-stress episode.  Gold borders mark the best method in each regime.  Standard KF values are shown as annotations (off-scale).  No single fixed-bandwidth EWMA dominates across all regimes.}
  \label{fig:finance_loglik}
\end{figure}

\begin{figure}[htbp]
  \centering
  \includegraphics[width=\columnwidth]{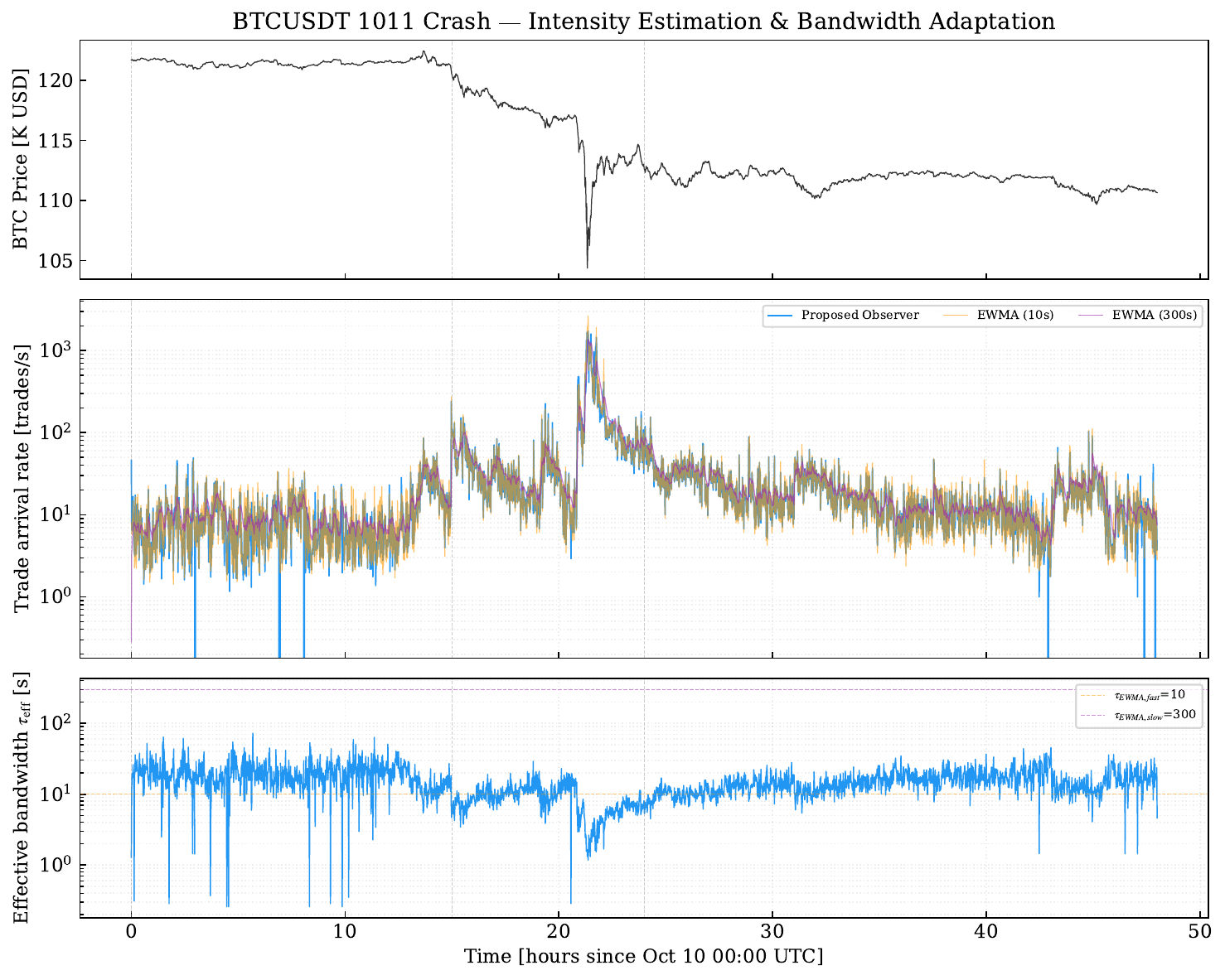}
  \caption{Adaptive intensity estimation and bandwidth contraction during the BTCUSDT market-stress episode.  Top: BTC price.  Middle: estimated trade arrival rate (log scale).  Bottom: effective bandwidth $\tau_{\mathrm{eff}}$ of the proposed observer.}
  \label{fig:finance_intensity}
\end{figure}

\Cref{tab:finance_loglik} lists the exact regime-wise predictive log-likelihood values.  \Cref{fig:finance_loglik} presents the same comparison as a bar chart, and \Cref{fig:finance_intensity} visualizes the observer's adaptive behavior during the BTCUSDT market-stress episode.  The effective bandwidth $\tau_{\mathrm{eff}}$ spans a 407-fold dynamic range, from 0.22\,s at the activity peak to 89.7\,s during overnight quiescence, governed entirely by the Riccati recursion.

\begin{table}[htbp]
\centering
\caption{Three-Regime Predictive Log-Likelihood on BTCUSDT (Oct~10--11, 2025).}
\label{tab:finance_loglik}
\begin{tabular}{lccc}
\toprule
Method & Pre-event & Event & Post-event \\
\midrule
Proposed observer & $-13.36$ & $\mathbf{-40.30}$ & $-14.54$ \\
EWMA (5\,s)       & $-12.86$ & $-40.53$           & $-14.83$ \\
EWMA (10\,s)      & $-12.42$ & $-40.55$           & $-14.38$ \\
EWMA (15\,s)      & $-12.27$ & $-40.98$           & $-14.27$ \\
EWMA (300\,s)     & $-12.61$ & $-53.49$           & $-14.74$ \\
Standard KF       & $-49.45$ & $-100.26$          & $-41.74$ \\
\bottomrule
\end{tabular}
\end{table}

\FloatBarrier

\section{Computational Profiling}
\label{sec:profiling}

The $\mathcal{O}(1)$ per-step complexity established in this paper is verified empirically through wall-clock timing experiments.  All timing experiments were conducted on an Intel Core i7-13620H running at 2.40\,GHz with 16\,GB RAM, using Python~3.9.25 on Windows~11.

Event sequences of increasing length $N \in \{10^3, 10^4, 10^5, 10^6\}$ are generated from a synthetic Hawkes process to preserve realistic clustering structure.  The proposed observer and the oracle offline adaptive KDE are executed on a single-threaded CPU, and total wall-clock time is recorded for each sequence length.  All timing experiments are repeated 20 times and the median execution time is reported to minimize the influence of system scheduling variability.

\begin{figure}[htbp]
  \centering
  \includegraphics[width=\columnwidth]{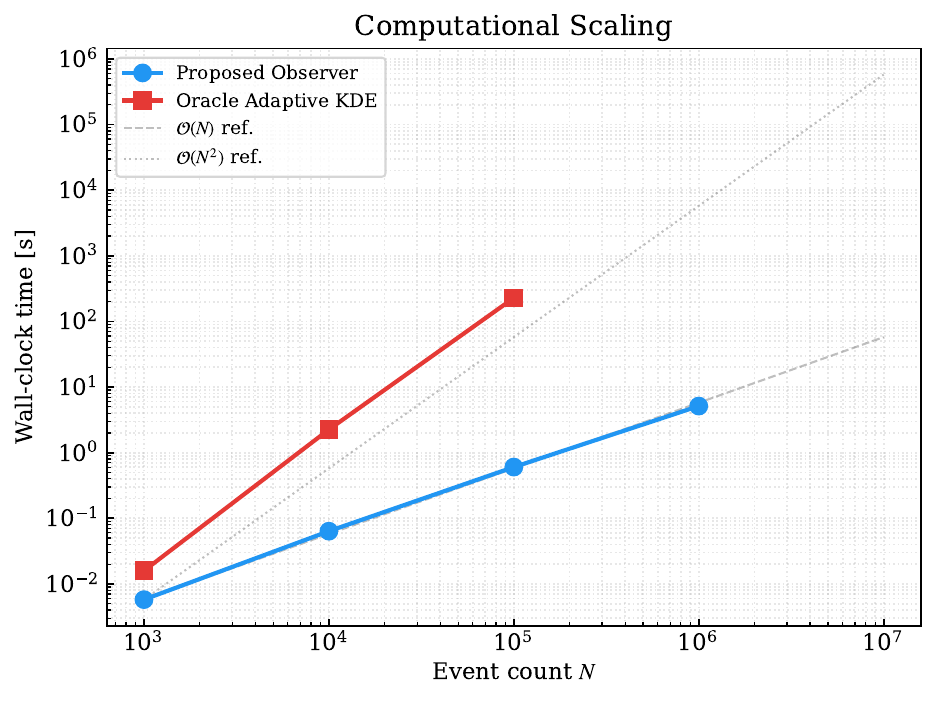}
  \caption{Per-step computational cost: $\mathcal{O}(1)$ observer versus $\mathcal{O}(N^2)$ oracle KDE.  The log-log plot shows that the observer's total time scales linearly with $N$, while the oracle KDE follows the expected quadratic trajectory.  The oracle KDE at $N = 10^6$ is not shown because its projected execution time exceeds 60 hours.}
  \label{fig:complexity_scaling}
\end{figure}

\begin{table}[htbp]
\centering
\caption{Wall-Clock Time and Per-Bin Cost Across Three Orders of Event Scale.}
\label{tab:timing}
\begin{tabular}{lcccc}
\toprule
Method & $N\!=\!10^3$ & $N\!=\!10^4$ & $N\!=\!10^5$ & $N\!=\!10^6$ \\
\midrule
Proposed observer & 0.006\,s & 0.064\,s & 0.604\,s & 5.133\,s \\
Oracle KDE        & 0.016\,s & 2.251\,s & 228.3\,s & ---      \\
\midrule
Observer $\mu$s/bin & 28.7 & 31.8 & 30.2 & 25.7 \\
\bottomrule
\end{tabular}
\end{table}

As shown in \cref{fig:complexity_scaling} and \cref{tab:timing}, the observer's per-bin cost remains between 25.7 and 31.8\,$\mu$s across event counts spanning three orders of magnitude ($N = 10^3$ to $10^6$), confirming the $\mathcal{O}(1)$ per-step complexity.  Total wall-clock time scales linearly with the number of bins: 0.006\,s at $N = 10^3$, rising to 5.13\,s at $N = 10^6$.

The oracle adaptive KDE, by contrast, exhibits clear $\mathcal{O}(N^2)$ scaling: from 0.016\,s at $N = 10^3$ to 228\,s at $N = 10^5$, a ratio of approximately 14{,}000 $\approx (100)^2$, matching the expected quadratic trajectory.  At $N = 10^6$ the KDE computation becomes infeasible within a practical time budget.  The observer therefore delivers a speedup of 2.7$\times$ at $N = 10^3$, growing to 378$\times$ at $N = 10^5$, with the gap widening quadratically beyond that scale.

\section{Discussion}
\label{sec:discussion}

The empirical evidence assembled in \cref{sec:synthetic,sec:applications} admits a single structural interpretation of the observer's behavior across heterogeneous regimes.  On Profile~A the fast EWMA amplifies Poisson noise during quasi-stationary segments while the slow EWMA incurs a 1.9\,s phase lag at step transitions, and on the BTCUSDT data the two time constants exchange rank between the event window and the post-event period (\cref{tab:profile_a_results,sec:app_finance}).  No fixed-bandwidth estimator can resolve this conflict across regimes, whereas the proposed observer adapts autonomously because the Riccati equation couples Poisson measurement noise with geometric process diffusion to yield the gain scaling $K_{ss} \propto \sqrt{\lambda}$ and the bandwidth law $\tau_{\text{eff}} \propto \lambda^{-1/2}$ (\cref{thm:sqrt_gain,cor:bandwidth}).  The bandwidth is thus endogenous to the noise axioms rather than supplied by external tuning, and the same mechanism predicts a shallower exponent $\tau_{\text{eff}}^{\text{CV}} \propto \lambda^{-1/4}$ for the constant-velocity extension (\cref{prop:cv_scaling}), because the velocity state absorbs the first-order intensity gradient and shifts the dominant residual from bias to curvature.  Numerical fits of $-0.505$ and $-0.251$ in \cref{fig:scaling_law_numerical} match both predictions to sub-percent accuracy, and the 11.3\% RMSE reduction from the scalar to the CV observer on Profile~A (\cref{tab:profile_a_results}) localizes the gain to the sinusoidal segments where the velocity state tracks a smooth gradient, recovering the classical $\alpha$-$\beta$ trade-off~\cite{Kalata1984}.

Three limitations circumscribe the present formulation and define concrete directions for future work.  The linear Kalman update relies on a Gaussian approximation to the binned Poisson count, which becomes progressively less accurate in very sparse regimes where $\lambda_k \Delta t$ approaches or falls below one event per bin and the distribution is dominated by zeros and ones; the neural results in \cref{sec:app_neuro} indicate that the observer remains useful at moderate count levels, but substantially sparser settings will likely require nonlinear extensions such as an extended or unscented Kalman filter, or sequential Monte Carlo updates with a Poisson observation likelihood, while offline analyses can invoke a Rauch-Tung-Striebel backward pass~\cite{Rauch1965} for non-causal smoothing.  A separate obstacle to scale is the univariate scope, which excludes the structured cross-channel dependencies present in high-density neural recordings and correlated order flows across venues, where a naive extension to dimension $d$ would incur $\mathcal{O}(d^3)$ cost per step and forfeit the constant-time advantage; block-diagonal covariance models and the data-driven Kalman variants of~\cite{Revach2022,Buchnik2024} offer a principled route that can reuse the endogenous noise parametrization as physically grounded inductive bias.  Exogenous modulation is likewise absent from the current formulation: the observer is driven entirely by the endogenous event history, so stimulus-response paradigms, news-driven trading surges, and seasonally modulated seismic activity require an exogenous control term $\mathbf{B}\mathbf{u}_k$ in the state-transition equation, an extension that is algebraically straightforward and leaves the Riccati-based bandwidth mechanism intact.

\section{Conclusion}
\label{sec:conclusion}

This paper establishes a formal duality between Kalman filtering and adaptive kernel smoothing for point process intensity estimation.  Under two physically motivated noise axioms, Poisson counting statistics dictating measurement uncertainty ($R \propto \lambda$) and geometric diffusion governing process volatility ($Q \propto \lambda^2$), the steady-state algebraic Riccati equation yields an effective smoothing bandwidth that scales as $\tau_{\text{eff}} \propto \lambda^{-1/2}$.  This result gives a causal, recursive, $\mathcal{O}(1)$ per-step realization of square-root adaptive bandwidth scaling and explains how that scaling emerges from the interaction between Poisson measurement noise and geometric process diffusion.

Empirical results support the theoretical predictions.  Controlled synthetic experiments confirm the predicted scaling exponents for both scalar and constant-velocity model orders and the geometric noise model's capacity to track self-exciting avalanches, while additional experiments verify the analytical divergence of reciprocal estimators and the dormant-state gating mechanism's suppression of phantom intensity leakage.  Two cross-domain applications, to motor-cortex spike trains and tick-level cryptocurrency trade data, indicate competitive goodness-of-fit relative to domain-tuned baselines under model-free diagnostics that require no access to ground-truth intensity.  Computational profiling yields an empirical per-step cost of approximately 30\,$\mu$s.

The framework opens several avenues for future investigation, including nonlinear filtering extensions for low-count regimes, multi-channel generalizations for high-dimensional neural and financial data, and the incorporation of exogenous covariates for stimulus-response modeling.  More broadly, the results demonstrate that embedding the statistical structure of the observed phenomenon directly into the noise parametrization of a state-space model can cause classical optimality results to emerge autonomously from the filtering dynamics, suggesting that similar duality principles may govern adaptive estimation in other counting and renewal processes.

\section*{Code Availability}
The implementation, synthetic experiments, and evaluation tools are publicly available at \url{https://github.com/chin1hanwu/CAIE-Kalman-Kernel-Duality-Code/tree/v0.1.0}.

\section*{Acknowledgments}
The authors used an AI language model (Claude Opus 4.6, Anthropic; accessed via Google Antigravity) as an interactive programming assistant to accelerate prototyping, debugging, and code refactoring during the experimental implementation.  All AI-assisted code was manually reviewed by the authors, validated through reproducible execution of the computational experiments, and verified with unit tests where applicable.  The authors take full responsibility for the manuscript content, analyses, and conclusions.

\bibliographystyle{unsrtnat}
\bibliography{refs}

\end{document}

%% file: figures/scaling_law.tex
\begin{figure}[t]
\centering
\begin{tikzpicture}
\begin{loglogaxis}[
    width=\columnwidth,
    height=0.7\columnwidth,
    xlabel={Intensity $\lambda$ (events/s)},
    ylabel={Effective bandwidth $\tau_{\text{eff}}$ (s)},
    xmin=0.1, xmax=1000,
    ymin=0.05, ymax=50,
    grid=both,
    grid style={line width=0.3pt, draw=gray!30},
    major grid style={line width=0.5pt, draw=gray!50},
    legend style={
        at={(0.5,-0.22)},
        anchor=north,
        font=\scriptsize,
        draw=gray!60,
        fill=white,
        fill opacity=0.9,
        text opacity=1,
        rounded corners=2pt,
        row sep=1pt,
        column sep=4pt,
        /tikz/every even column/.append style={column sep=8pt},
    },
    legend columns=2,
    tick label style={font=\scriptsize},
    label style={font=\footnotesize},
    every axis plot/.append style={thick},
]


\addplot[
    domain=0.1:1000,
    samples=200,
    color=black!50,
    dashed,
    line width=1.0pt,
] {10 * x^(-0.5)};
\addlegendentry{$\tau \propto \lambda^{-1/2}$ (Abramson)}

\addplot[
    domain=0.1:1000,
    samples=200,
    color=blue!80!black,
    solid,
    line width=1.5pt,
] {
    -0.1 / ln(1 - (
        (0.01*x^2 + sqrt(0.0001*x^4 + 4*0.01*x^2*1*x/0.1)) / 2
    ) / (
        (0.01*x^2 + sqrt(0.0001*x^4 + 4*0.01*x^2*1*x/0.1)) / 2 + 1*x/0.1
    ))
};
\addlegendentry{1D scalar (exact DARE)}

\addplot[
    domain=0.1:1000,
    samples=200,
    color=red!65!black,
    dotted,
    line width=1.0pt,
] {sqrt(1*0.1/(0.01*x))};
\addlegendentry{1D asymptotic: $\sqrt{\phi\Delta t/(q\lambda)}$}


\addplot[
    domain=0.1:1000,
    samples=200,
    color=black!50,
    dash dot,
    line width=1.0pt,
] {10 * x^(-0.25)};
\addlegendentry{$\tau \propto \lambda^{-1/4}$ (reference)}

\addplot[
    domain=0.1:1000,
    samples=200,
    color=green!55!black,
    solid,
    line width=1.5pt,
] {(25)^(0.25) * x^(-0.25)};
\addlegendentry{2D CV (CARE)}

\draw[red!60!black, line width=1.2pt, dashed]
    (axis cs:100,0.05) -- (axis cs:100,50)
    node[pos=0.82, right, font=\scriptsize, text=red!60!black,
         fill=white, fill opacity=0.8, text opacity=1,
         inner sep=1.5pt, rounded corners=1pt] {$\rho = 0.1$};

\end{loglogaxis}
\end{tikzpicture}
\caption{Effective smoothing bandwidth $\tau_{\text{eff}}$ as a function of local
intensity $\lambda$ on a log-log scale.
The 1D scalar model (blue solid) closely tracks the Abramson
reference $\tau \propto \lambda^{-1/2}$ (gray dashed) and the closed-form
asymptotic approximation $\tau = \sqrt{\phi\Delta t/(q\lambda)}$ (red dotted)
throughout the practical operating regime $\rho < 0.1$.
The 2D Constant Velocity model (green solid), computed from the continuous
algebraic Riccati equation, follows the shallower trajectory
$\tau \propto \lambda^{-1/4}$ (gray dash-dot),
reflecting the reduced approximation bias afforded by inertial trend tracking
(\cref{prop:cv_scaling}).
Parameters: $q = 0.01$, $\phi = 1$, $\Delta t = 0.1$\,s.}
\label{fig:scaling_law}
\end{figure}